\documentclass[aip,apl,graphicx,reprint]{revtex4-1}

\usepackage{multirow}
\usepackage{siunitx}
\usepackage{graphicx}
\newcommand{\changed}[1]{#1}
\newcommand{\ket}[1]{{|{#1}\rangle}}

\usepackage{hyperref}

\begin{document}


\title{Organic molecules as single-photon sources}

\author{A. Shkarin}
\email{alexey.shkarin@mpl.mpg.de}
\affiliation{Max Planck Institute for the Science of Light, D-91058 Erlangen, Germany}
\author{S. G{\"o}tzinger}
\email{stephan.goetzinger@mpl.mpg.de}
\affiliation{Max Planck Institute for the Science of Light, D-91058 Erlangen, Germany}
\affiliation{Department of Physics, Friedrich Alexander University Erlangen-Nuremberg, D-91058 Erlangen, Germany}
\affiliation{Graduate School in Advanced Optical Technologies (SAOT), Friedrich Alexander University Erlangen-Nuremberg, D-91052 Erlangen, Germany}

\date{\today}

\begin{abstract}
The development of single-photon sources has been nothing but rapid in recent years, with quantum emitter-based systems showing especially impressive progress. In this article, we give an overview of the developments in single-photon sources based on single molecules. We will introduce polycyclic hydrocarbons as the most commonly used emitter systems for the realization of an organic solid-state single-photon source. At cryogenic temperatures this special class of fluorescent molecules demonstrates remarkable optical properties such as negligible dephasing, indefinite photostability, and high photon rates, which make them attractive as fundamental building blocks in emerging quantum technologies. To better understand the general properties and limitations of these molecules, we discuss sample preparation and relevant emitter parameters such as absorption and emission spectra, lifetime, and dephasing. We will also give an overview of light extraction strategies as a crucial part of a single-photon source. Finally, we conclude with a look into the future, displaying current challenges and possible solutions.
\end{abstract}


\maketitle

\section{A brief history of single molecules in quantum optics}
Single molecules were the first emitters to be detected in a solid-state system \cite{Moerner1989}. In 1989 Moerner and Kador published a seminal paper investigating single pentacene molecules in a \textit{para}-terphenyl organic host crystal. In their experiment, molecules were detected via an absorption measurement at cryogenic temperatures, which required a double modulation technique due to the weak absorption signal of a single molecule. Only one year later, Bernard and Orrit established fluorescence excitation spectroscopy as a new method to detect single molecules \cite{Orrit1990}, and in the same year, single molecules were also observed in solution at room temperature\cite{BrooksShera1990}. Detecting any kind of single emitter via fluorescence has since become the method of choice due to superior signal-to-background ratio and ease of use \cite{Orrit2008}. In fact, single-molecule detection in solution or on a surface has become one of the work-horses in life sciences \cite{Joo2008, Shashkova2017}. Meanwhile, extinction measurements have continued to advance, now demonstrating extinction levels exceeding \SI{10}{\percent} attenuation of light by a single molecule\cite{Gerhardt2007, Wrigge2008, Wrigge2008b} and even allowing for the detection of single emitter absorption at room temperature\cite{Kukura2009, Kukura2010}.

Experiments towards the use of single molecules as single-photon sources have been performed as early as 1992, when photon anti-bunching was reported \cite{Basche1992}, confirming the non-classical properties of the light emitted by a molecule. Other studies in the mid 1990s include the implementation of Stark tuning \cite{Wild1992,Orrit1992} and exploration of photon bunching to investigate the internal photophysics of a single molecule, like the intersystem crossing rate and triplet state lifetime \cite{Bernard1993}.

By the year 2000, triggered single-photon emission has been reported in several solid-state material systems, including molecules, quantum dots, and NV centers \cite{Brunel1999a, Lounis2000, Michler2000, Kurtsiefer2000}. Since then, the field has flourished, and numerous single-photon sources based on various material systems have been demonstrated. The fluorescent molecules investigated for that role mostly come from the group of polycyclic aromatic hydrocarbons (PAH), as they are the best at fulfilling the stringent requirements for single-photon sources, showing indefinite photostability and high photon indistinguishability at cryogenic temperatures. The latter is a key requirement for quantum information processing and has thus become a figure-of-merit in the context of single-photon sources.

The first indistinguishable photons emission by a single molecule was demonstrated in 2005 using terrylenediimide (TDI)\cite{Kiraz2005}. Following that, two-photon interference of photons emitted by two remote dibenzanthanthrene (DBATT) molecules was reported as a conceptual proof that independent molecules can deliver high-quality indistinguishable photons \cite{Lettow2010}. Since these early experiments, groups around the world have advanced the engineering of high-performance single-molecule single-photon sources, as has been highlighted in several recent reviews \cite{Moerner2004,Toninelli2021,GaitherGanim2022}.

This review is organized as follows. We start by introducing molecular quantum emitters and their basic underlying photophysics. Afterwards, we describe the emitter properties which are related to single-photon generation and present their values for the most common systems. Next, we discuss photon collection methods that have been employed with single-molecule emitters. After that, we present the current state of the art in the most relevant single-photon source parameters. Finally, we close by listing outstanding challenges and possible future directions.

\section{Molecular quantum emitter}
\label{sec:overview}

\begin{figure*}
\includegraphics[width=16.5cm]{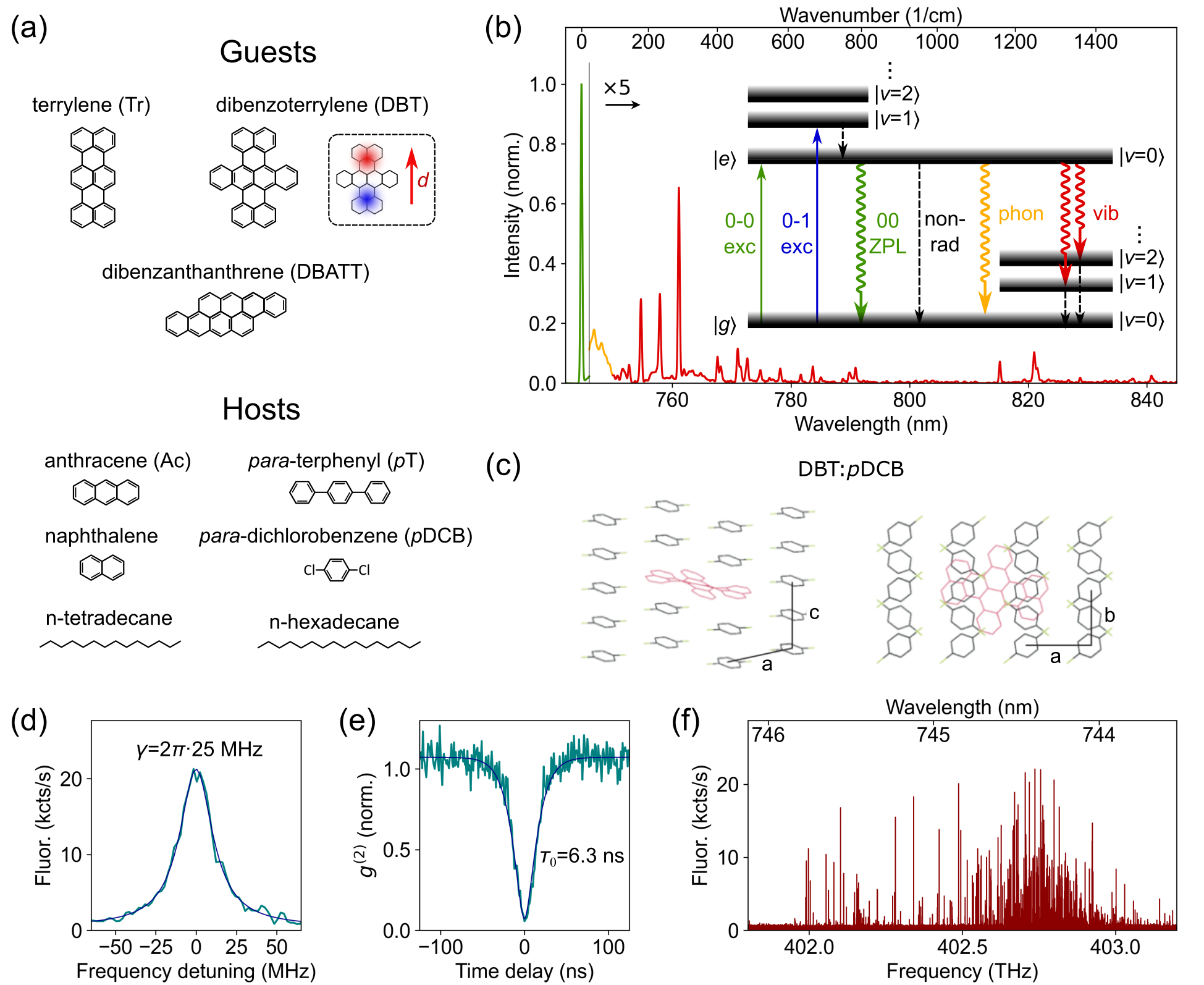}
\caption{\label{fig:overview} Basic properties of molecular quantum emitters. (a) Molecular structures of the most common guest and host species. The inset next to the structure of dibenzoterrylene (DBT) illustrates the electronic density associated with its $S_0\rightarrow S_1$ transition\cite{Sadeq2018}. (b) Emission spectrum of DBT in \textit{p}DCB highlighting its 0-0 zero phonon line (00ZPL; green), phonon wing of the zero phonon line (yellow) and vibronic transitions (red). The inset shows a Jablonski diagram of a single molecule. (c) Illustrations of the insertion site of DBT in \textit{p}DCB\cite{Zirkelbach2022} from two different direction. (d) 00ZPL excitation spectrum of a single DBT molecule in \textit{p}DCB under weak excitation. The indicated linewidth $\gamma$ (full width at half maximum) is extracted by fitting a Lorentzian expression $F(\omega)=F_0\frac{(\gamma/2)^2}{(\omega-\omega_0)^2+(\gamma/2)^2}$, where $F_0$ is the maximal fluorescence signal and $\omega-\omega_0$ is the excitation laser detuning. (e) Second order autocorrelation function of the red-shifted emission from the same molecule under resonant excitation. (f) Excitation spectrum for DBT molecules in \textit{p}DCB over a range of about \SI{1}{THz} showing several hundreds of individual molecule resonances. Images in (c) are reproduced from Zirkelbach \textit{et al.},\cite{Zirkelbach2022} Journal of Chemical Physics \textbf{156}, 104301 (2022). Copyright 2022 Author(s), licensed under a Creative Commons Attribution (CC BY 4.0) license.}
\end{figure*}

\subsection{Electronic properties}

The molecules discussed in this review all belong to the class of PAHs, with some examples shown in Fig. \ref{fig:overview}a. They consist of only two chemical elements: carbon, forming the backbone of joined benzene (i.e., aromatic) rings, and hydrogen, which terminates the structure on the perimeter (not shown explicitly in Fig. \ref{fig:overview}a). The fundamental pattern is reminiscent of that of graphene, and, indeed, there are similarities in the fundamental optical properties and some computational methods used to calculate the optical response\cite{Sadeq2018}.

The origin of the optical transition in such molecules can be explained in the same terms as used for individual atoms. Just like in atoms, the electrons in molecules occupy a ladder of well-defined states called molecular orbitals, which are delocalized over multiple atoms. If a pair of these orbitals has different parity, then promotion of the electron from one to the other will be associated with a non-zero electrical dipole matrix element, resulting in an optically active transition. The lowest-energy transition, which is considered most often, corresponds to a single electron moving from the highest occupied molecular orbital (HOMO) to the lowest unoccupied molecular orbital (LUMO).

When considering the optical activity of a given transition, it is important to take spin properties of the corresponding states into account. The electronic ground state of the commonly used PAHs is spin-singlet, i.e., the total electron spin is zero. If the excited state has a different spin configuration, e.g., spin triplet, then the transition must rely on the relatively weak spin-orbit coupling to flip the electron spin, which drastically reduces its optical strength. Therefore, even though the triplet excited states have lower energy than the singlet, they are rarely addressed directly in PAH experiments. Nevertheless, their presence becomes important when emission intermittence is concerned, as discussed in section \ref{sec:emitter_parameters}.

Considering the above, the electronic structure is simplified to the two lowest spin-singlet states $S_0$ (i.e., the ground states of the system) and $S_1$, which are simply referred to as the electronic ground and excited states ($\ket{g}$ and $\ket{e}$ in the inset of Fig. \ref{fig:overview}b). An example of a transition between two such states is shown in the inset of Fig. \ref{fig:overview}a, which plots the electron density of the transition matrix element.

\subsection{Host material}

A solid-state material that contains the molecules is often called host material or host matrix (chemical structures of common host materials are shown in Fig. \ref{fig:overview}a), while the embedded optically active molecules are referred to as guests. Since the interaction between the host and the guest molecules is weak, the emitters preserve their molecular orbitals and, correspondingly, their energy level structure when inserted into the host.

Unlike most other solid-state emitters (e.g., color centers or epitaxial quantum dots), molecular emitters do not inherently rely on the host material, and they can still be optically addressed in a liquid or in the gas phase. However, the solid host material serves several purposes and is in fact a requirement for most quantum optical experiments. First, it immobilizes the molecule; one can therefore work with the same molecule for prolonged periods of time, reaching several years at cryogenic temperatures. Second, the matrix also restricts the rotational motion, which removes rotational transitions from the spectrum, greatly simplifying it. Third, it protects the molecule from outside contaminants, which is especially important for room temperature applications. Finally, interaction with the matrix results in fast dissipation of the vibrational modes (discussed in more detail in the next subsection), ensuring that the molecule is almost always in the vibrational ground state. This avoids trapping of the system in higher-lying vibrational states, which changes the optical transition frequency and brings the molecule out of resonance with the excitation laser.

In order to achieve good emitter properties, the host material has to satisfy several requirements. First, it should be technologically compatible with the process of inserting guest molecules. For example, if the molecules are introduced in the liquid host phase, the temperature should be low enough so that the guests do not decompose or oxidize. For this reason, the most commonly used materials are organic crystals with relatively low melting or sublimation temperatures below \SI{300}{\degree C}. Second, the host and guest should be geometrically compatible, i.e., there should be a way for individual guest molecules to insert themselves into the host structure. In crystalline hosts, due to their ordered structure, this produces well-defined molecular configurations referred to as insertion sites (two examples are demonstrated in Fig. \ref{fig:overview}c). This requirement is the reason why most crystalline host materials are molecular crystals with smaller molecules compared to the guest, as this gives more opportunities to produce space for the guest by removing several of the host molecules\cite{Nicolet2007b}. Finally, the host should be energetically compatible with the guest. This means that its lowest electronic excitations (including the triplet state\cite{Nicolet2006}) should have higher energy than the guest molecule's transition to avoid energy transfer from guest to host, which would reduce the quantum yield of the guest's optical transition.

With the correct choice of the material and low enough temperatures (typically around 2-\SI{4}{K}), the host provides an ordered and stable environment for the guest, resulting in negligible dynamic effects on the guest's resonance frequency. As a consequence, guest emitters display perfect spectral stability and Fourier-limited optical transitions \cite{Nicolet2007,Rezai2018,Ren2022,Musavinezhad2023,Huang2025}, meaning that the host-induced dephasing is much lower than the natural transition linewidth (as exemplified in Fig. \ref{fig:overview}d,e).

At the same time, the presence of polar or polarizable host molecules still leads to a significant static change of the guest molecule's optical properties. First, the increased photonic density of states enhances the radiative decay rate\cite{Rikken1995}, resulting in shorter-lived excited states and broader Fourier-limited linewidth. Second, it can affect the energies of the excited states, producing a so-called solvent shift in the optical transition frequency\cite{Suppan1990}. Given the unavoidable inhomogeneity of the host material, this shift varies slightly for different guest molecules, which produces a distribution of the resonance frequencies referred to as inhomogeneous broadening (IHB). An example is shown in Fig. \ref{fig:overview}f, which displays several hundred of individual guest molecule resonances, each about \SI{30}{MHz} wide, spread over a range of about \SI{1}{THz}. This figure also exemplifies the high achievable density of molecular emitters, as all of these observed resonances come from the confocal excitation volume of about \SI{1}{\micro m^3}.

The host materials commonly used with PAHs can be split into two classes depending on their degree of local order. The first are the well-ordered molecular crystals such as anthracene (Ac), \textit{para}-dichlorobenzene (\textit{p}DCB), \textit{para}-terphenyl (\textit{p}T), and naphthalene (all shown in Fig. \ref{fig:overview}a). They provide the most uniform and stable environment for the guest molecules, which results in low static and dynamic disorder, manifesting as a narrow IHB, high spectral stability, and low dephasing. On the other hand, these materials have stricter demands on the geometric matching, as the guest molecules essentially represent impurities for the host crystal, so they tend to get expelled during preparation. Also in this class are Shpol'skii matrices such as \textit{n}-tetradecane or \textit{n}-hexadecane (long chains in Fig. \ref{fig:overview}a), which are liquid at room temperature and solidify upon rapid cryogenic cooling, forming a polycrystalline structure. The second class of materials are polymer amorphous hosts such as poly(methyl methacrylate) (PMMA) or polyethylene (PE). They form a very disordered environment (although some reports suggest that certain kinds of PE have a degree of local crystallinity\cite{Orrit1992,Wirtz2006,Rattenbacher2023}) with a complicated energy landscape, resulting in a large IHB and poor spectral stability of the emitters. Nevertheless, their higher chemical and thermal stability and good optical quality make them potentially attractive in hybrid systems \cite{Rattenbacher2023}, where relatively volatile molecular crystal systems are harder to accommodate.

\subsection{Vibrations and phonons}

Compared to individual atoms, molecules possess extra degrees of freedom associated with the relative motion of the constituent nuclei. A molecule consisting of $N$ atoms has, in vacuum, $3N$ degrees of freedom, with three corresponding to translations and three to rotations (ignoring purely linear molecules such as carbon dioxide), leaving the remaining $3N-6$ as vibrations. When embedded into a solid-state material, translation and rotation are suppressed; as a result, all $3N$ degrees of freedom are associated with different vibrational modes, though the lowest frequency modes can still be reminiscent of restricted translations and rotations of the whole guest molecule. These low-frequency vibrational modes can mix with the host matrix vibrations, which leads to their delocalization and gives rise to the so-called pseudolocal modes\cite{Fleischhauer1992}.

The importance of the vibrational modes comes from their optical activity. Since the electronic density distribution is different in $\ket{g}$ and $\ket{e}$, the interaction potential of the nuclei composing the molecule is also slightly different, resulting in different equilibrium positions. This difference, in turn, implies that the full transition matrix element, which involves both electronic and nuclear (i.e., vibrational) degrees of freedom, will be non-zero even when the vibrational state changes during the transition. When applied to emission, this results in the creation of one or more vibrational quanta; as the total energy should be conserved, the energy of the emitted photon gets lowered, resulting in the red-shifted fluorescence (red part of the spectrum in Fig. \ref{fig:overview}b and red wavy arrows in its inset). Similarly, vibrations can also be created during the excitation, which produces additional excitation transitions with higher energy (blue line in the inset of Fig. \ref{fig:overview}b). Collectively, such optical transitions which involve a change of the vibrational state are called vibronic transitions. The main electronic transition between the vibrational ground states of $\ket{e}$ and $\ket{g}$ (green peak in Fig. \ref{fig:overview}b and green line in its inset) is often referred to as 0-0 zero-phonon line (00ZPL). As it does not involve any short-lived vibrations, it is the narrowest and the most commonly addressed optical transition in molecular emitters.

The frequencies of the vibrational modes typically span the range from about \SI{10}{cm^{-1}} (\SI{300}{GHz}) for the lowest-frequency large-scale vibrations to about \SI{3000}{cm^{-1}} (\SI{90}{THz}) for the highest frequency C-H bond stretching vibrational modes\cite{Biakowska2017,Greiner2020}. At cryogenic temperatures below \SI{10}{K} (frequency of \SI{200}{GHz}) all of these modes are predominantly in their ground state at equilibrium. Furthermore, in all molecules considered in this review, the vibrational lifetime is two to three orders of magnitude shorter than that of the electronic transition. As a result, whenever the molecular vibrations are excited, they are very likely to decay before the molecule changes its electronic state. This behavior is referred to as Kasha's rule\cite{Kasha1950}, which states that at any given moment the emitter is much more likely to be in the lowest energy vibrational state of the corresponding electronic manifold (i.e., the lowest states among the ones marked $\ket{g}$ and $\ket{e}$). As a consequence, only transitions originating from these states are considered when describing the emitter behavior under normal circumstances (although strong enough excitation can result in stimulated transitions which violate this rule\cite{Zirkelbach2023}).

The vibronic transitions into $v>0$ manifold of the electronically excited state are frequently exploited for excitation in SPS applications, which is commonly referred to as 0-1 excitation scheme (blue line in the inset of Fig. \ref{fig:overview}b). Due to their relatively narrow linewidth of several \SI{}{GHz} to several tens of \SI{}{GHz}\cite{Zirkelbach2022}, they can be used to efficiently drive the molecule to its excited state. Thanks to Kasha's rule, the emission still occurs from the vibrational ground state of $\ket{e}$, so it has a significant fraction of the narrow 00ZPL producing indistinguishable photons. At the same time, the fast vibrational state decay allows for the excited state inversion, which further boosts the photon emission rate. Combined with the fact, that simple spectral filtering allows for easy separation of the excitation laser from the emitted single photons, this excitation scheme is the most commonly used for generating indistinguishable photons from a single molecule\cite{Lettow2010,Rezai2018,Lombardi2021,Duquennoy2022,Schofield2022b}.

In addition to intramolecular vibrations described above, there are also delocalized vibrations of the host material, i.e., phonons. These mainly have two effects on the molecular emitters stemming from two kinds of interaction. The first case originating from a linear interaction is analogous to the effect of the intramolecular vibrations: the phonons get created during optical transitions, giving rise to additional red-shifted emission (yellow part of the spectrum in Fig. \ref{fig:overview}b and yellow wavy lines in its inset) and blue-shifted absorption. Since the phonons form a continuum rather than discrete modes, this effect produces relatively broad spectral features on the scale of several \SI{}{THz}, which are often called phonon wings. The second phonon effect, stemming from the quadratic electron-phonon interaction, gives rise to temperature-dependent dephasing of the optical transitions, as described in the section \ref{sec:emitter_parameters}.

\section{Preparation techniques}
\label{sec:preparation}

Molecules as quantum emitters have certain advantages, such as their availability, simplicity of handling, and high achievable concentration. This is in large part due to their preparation relying on standard chemical synthesis methods, producing macroscopic quantities of emitters. The creation of the sample then requires little more than dissolving these molecules in an appropriate host material. This straightforward approach allows for a large variety of host materials and concentrations, from more than \SI{100}{\micro m^{-3}} down to a single molecule per observation volume\cite{Shkarin2021,Zirkelbach2022,Rattenbacher2024}. It is important to note, that molecules are well-dispersed within the host material for all of the reported preparation techniques and do not show signs of clustering on nanometer scales. Hence, isolated peaks in the ZPL excitation spectrum (such as the ones shown in Fig. \ref{fig:overview}d and f) can almost always be attributed to a single emitter, as is frequently confirmed through second-order intensity autocorrelation measurements (e.g., Fig. \ref{fig:overview}e).

Several methods of distributing molecules in a host material exist, which differently balance the optical emitter quality, simplicity of preparation, and ease of integration. On one end of the spectrum is the co-sublimation method, which involves heating the host and the guest material together and forming a vapor, from which single crystals are then grown, typically with dimensions on the scale of 10's of \SI{}{\micro m} to \SI{}{mm}. This method generally leads to the highest host matrix quality, which manifests as the narrowest IHB and the lowest spectral diffusion and dephasing. On the other hand, this method is fairly complicated, requiring careful tuning of growth parameters, and the resulting fully formed crystals are challenging to integrate into photonic structures due to their large size. Nevertheless, there have been successful demonstrations of interfacing them with optical devices presenting a large flat surface such as solid immersion lenses\cite{Trebbia2009}, Fabry-Perot cavities\cite{Wang2019}, and the top surface of nanophotonic waveguides\cite{Ren2022,Huang2025} and photonic crystal cavities\cite{Lange2025}.

Some other methods allow for more flexibility in matching the crystal to the surrounding geometry. One possibility is crystallization from a melt, where guest molecules are added in their powder form to a molten host material, which is subsequently cooled down below the crystallization temperature. This approach is commonly used where the host material needs to directly interface with other surfaces, such as nanophotonic circuits\cite{Turschmann2017,Boissier2021} or collection optics\cite{Wrigge2008,Siyushev2014,Zirkelbach2022}. To provide a well-defined sample geometry and inhibit host matrix sublimation, the molten material is often confined to a narrow channel before crystallization. This makes this technique well-suited for more volatile host materials with low melting temperatures such as naphthalene\cite{Jelezko1996,Jelezko1997,Gmeiner2016} or \textit{p}DCB\cite{Verhart2016}. The resulting sample is usually polycrystalline with crystal domains on the order of 10's to 100's of \SI{}{\micro m} and is typically somewhat stressed due to the confining geometry\cite{Gmeiner2016}, which might result in a larger IHB and overall spectral shift. A related approach uses Shpol'skii matrices such as \textit{n}-hexadecane\cite{Boiron1996} or \textit{n}-tetradecane\cite{Bloess2001}, which are liquid at room temperature but solidify during shock-freezing when the sample is cooled. In addition, some matrices such as Ac and \textit{p}T allow for thin film preparation\cite{Toninelli2010,Pfab2004,Checcucci2016,Lombardi2018}, where both the host and the guest material are dissolved in a common solvent and then spin coated onto a glass slide. This produces host crystals which are uniform and very thin, down to several 10's of \SI{}{nm}.

Another set of host preparation techniques focuses on the generation of very small molecule-doped crystals, often with sub-micron dimensions. The two most prominent techniques here are precipitation\cite{Pazzagli2018} and nanoprinting\cite{Hail2019,Musavinezhad2024}. In the first, host and guest materials dissolved in a water-soluble organic solvent are injected into sonicated water; as the solvents get dissolved in the water the host solution becomes oversaturated, which causes precipitation of doped host material nanocrystals. The resulting nanocrystals dispersed throughout the solution can then be introduced into the desired location and coated with a protection layer such as poly(vinyl alcohol) (PVA) to prevent their sublimation. While the nanocrystals are dispersed randomly throughout the solution, one could place them deterministically by pick and place techniques, which have been used in the past with gold nanoparticles and nanodiamonds\cite{Schell2011}. This method is relatively straightforward, produces crystals in large quantities, and is compatible with 3D-printed polymeric photonic devices\cite{Colautti2020b}. The other nanocrystal preparation method, nano-printing, involves applying electric voltage to a precisely positioned micron-sized nozzle, which ejects the solution of host and guest molecules onto a substrate. As the solvent evaporates, the doped host crystal is formed; despite the crystal dimensions being as small as $\SI{500}{nm}\times \SI{500}{nm} \times \SI{100}{nm}$, they were demonstrated to contain about 20 molecules\cite{Musavinezhad2024}. When relatively volatile host matrices, like Ac, are used, nanocrystals are often protected from sublimation by spin-coating a polymer layer similar to the precipitation approach. The nanoprinting technique allows for deterministic placement of nanocrystals with sub-\SI{}{\micro m} precision at the expense of a more complicated preparation setup and lower throughput, as the crystals are formed individually. In both nanocrystal methods the molecules generally show higher degree of dephasing and spectral diffusion compared to the more ``bulk'' methods such as co-sublimation or crystallization, even with the same guest and host material. This can be explained by a somewhat lower crystal quality of the nanocrystals and their smaller size, which necessarily means that the molecules are closer to the surface and therefore experience a more dynamic and unstable environment.

Somewhat separate from the preparation techniques described above is an approach which allows for the introduction of guest molecules into an already formed host material, as has been demonstrated with a PE host \cite{Wirtz2006,Rattenbacher2023}. It involves soaking the prepared sample (usually a thin film) in a solution of guest molecules dissolved in chloroform for a prolonged period of time, which allows the fluorescent molecules to diffuse into the PE. This is a promising approach to place emitters directly into nanophotonic structures, as it can be used as a last step in preparation of samples fabricated using techniques such as electron beam lithography or plasma etching, which might otherwise cause damage to dye molecules.

Similarly, some recent effort has been directed at combining organic molecules with hexagonal boron nitride (hBN)\cite{Neumann2023,Farooqui2025}, which has already been demonstrated to be compatible with the fabrication of photonic structures\cite{Froch2021,Spencer2023,Gerard2023,Schaeper2024}. While cryogenic studies are still lacking, these works chart a viable route towards molecular emitter integration with a highly stable and technologically attractive material.

Finally, a conceptually new method has been demonstrated recently, which involves deposition of molecules on the surface of a material rather than inserting them into the bulk. So far, two material platforms have been reported. The first uses terrylene (Tr) on the surface of hBN\cite{Han2021,Smit2022,DeHaas2025}, which offers excellent lattice matching and, as a consequence, well defined ``surface sites''. This system has already demonstrated a degree of dephasing and photostability comparable to the polymer matrices, with the observed dephasing attributed to the residual contaminants on the surface. The second work studies dibenzoterrylene (DBT) on the surface of anthracene\cite{Mirzaei2025}. While this combination has been used extensively in the past with molecules embedded into the crystal, this is the first report on surface emitters in this material system. Thanks to the combination of \textit{in-situ} guest deposition and host crystal self-cleaning through sublimation, this approach has delivered lifetime-limited linewidths, thus demonstrating the potential for high-quality surface-bound quantum emitters.

In all of the preparation methods discussed above, it is important to account for the impurities in the materials used for fabrication. These typically have two effects. First, at sufficiently high concentrations they can negatively affect the quality of the host matrix, resulting in extra dephasing or spectral diffusion. Here the purity of the host materials and solvents is more important, as they comprise the overwhelming majority (more than $99.9\%$ even for the highest reported guest doping concentration) of the input material. For this reason, high-purity chemicals are usually employed, and additional zone refinement steps are applied to the host material whenever appropriate. The second effect arises at high enough impurity concentration in the guest molecules, as this effectively ``dilutes'' them, resulting in the lower final density of fluorescent molecules. Nevertheless, this effect can be straightforwardly compensated by adjusting the material weight ratios.

\section{Emitter parameters}
\label{sec:emitter_parameters}

In this section we discuss various properties of molecular quantum emitters and their effect on the performance as a single-photon source (SPS). The most important of these parameters in the commonly used material systems are summarized in Table \ref{tab:material_systems}.

In the following discussion, we will categorize SPS based on their applications, which will determine the emitter requirements. First, we will distinguish between continuous-wave (CW) and on-demand pulsed sources. In the first case, the emitter is typically excited with a continuous laser drive and, as a result, the photons are continuously emitted at random intervals. Here, the single-photon character of the emission is expressed as a reduced probability of more than one photon being detected within a short enough time span whose scale is given by the excited state lifetime of the emitter. This property is uncovered by examining the second order autocorrelation function $g^{(2)}(\tau)$ (typically measured using a Hanbury Brown and Twiss setup), which shows a characteristic dip at zero time delay, i.e., $g^{(2)}(0)<0.5$ for a single dominant emitter (see Fig. \ref{fig:overview}e). As an alternative to the CW operation, one can use pulsed excitation to produce distinct single-photon pulses as long as the time between the excitation pulses is significantly longer than the excited state lifetime; in this case, the fundamental requirement is for each output pulse to contain at most one photon. As the single photon timing is more defined in this case, this method is generally preferred, although some metrology applications can be satisfied with CW sources\cite{Lombardi2020}. In terms of photon production, CW sources are typically described using the maximal achievable single photon rate, while pulsed sources are characterized via the maximal pulse repetition rate and the probability for an excitation pulse to produce a detectable photon.

The second distinction we will make is between sources of distinguishable and indistinguishable photons. In the first case, the only essential requirement is high single-photon purity, i.e., low probability of generating more than a single photon. Such sources are typically used in quantum cryptography, metrology, or sensing, where it is important to have a steady source of single photons, but these photons are completely independent and do not interact or interfere with each other. The second case adds an additional requirement on consecutive photons being indistinguishable, which requires their polarization, spatial mode, frequency, and time envelope to all be identical. Such indistinguishability is necessary in applications involving multiphoton interference, such as linear optics quantum computing\cite{Knill2001}. These sources are much harder to realize, particularly in regards to spectral matching, as any perturbation of the emitter frequency would reduce the photon indistinguishability.

\begin{table*}[t]
    \caption{\label{tab:material_systems}Material systems and their basic parameters}
    \begin{ruledtabular}
    \begin{tabular}{l|p{0.12\linewidth}|p{0.1\linewidth}|p{0.09\linewidth}|p{0.08\linewidth}|p{0.15\linewidth}|p{0.09\linewidth}|p{0.09\linewidth}}
        Guest:host & Preparation & $\lambda$, nm & IHB, nm & $\tau_0$, ns & $\gamma/2\pi$, MHz & $\eta$, \% & $\alpha_\mathrm{ZPL}$, \% \\\hline\hline
        DBT:Ac & co-sublimation & 785\cite{Nicolet2007,Wei2020,Ren2022} & 0.03\cite{Nicolet2007}, 2\cite{Ren2022} & 6\cite{Nicolet2007}, 4\cite{Ren2022,Huang2025} & 25-40\cite{Nicolet2007}, 40\cite{Ren2022}, 36\cite{Huang2025} & & 33\cite{Trebbia2009}, 46\cite{Schofield2022}\footnotemark[1] \\
        DBT:Ac & precipitation & 785\cite{Pazzagli2018,Lange2024} & 0.2\cite{Pazzagli2018} & 4\cite{Pazzagli2018}, 4.5\cite{Lange2024} & 40-80\cite{Pazzagli2018}, 30\cite{Lange2024} & &\\
        DBT:Ac & printing & 782\cite{Musavinezhad2024} & 2\cite{Musavinezhad2024},10\cite{Musavinezhad2024} & & 40-100\cite{Musavinezhad2024} & &\\
        DBT:Ac & melting & 790\cite{Turschmann2017} &  &  & 30\cite{Turschmann2017} & &\\
        DBT:PE & infusion & 740\cite{Rattenbacher2023} & 20\cite{Rattenbacher2023} & & 20-100\cite{Rattenbacher2023} & &\\
        DBT:\textit{p}DCB & melting & 745\cite{Verhart2016,Zirkelbach2022} & 1\cite{Musavinezhad2023} & 6.3\cite{Musavinezhad2023} & 25-30\cite{Verhart2016}, 23\cite{Zirkelbach2022}, 25\cite{Musavinezhad2023} & 70\cite{Musavinezhad2023}, 35\cite{Erker2022}\footnotemark[2] & 44\cite{Verhart2016}\footnotemark[1]\\
        DBT:naphthalene & melting & 758\cite{Jelezko1996}, 770\cite{Faez2014} & $>$3\cite{Faez2014} & 5.3\cite{Faez2014} & 25-35\cite{Jelezko1996}, 30\cite{Faez2014} & &\\
        DBT:\textit{p}T & precipitation & 770\cite{Schofield2022} & 3\cite{Schofield2022} & 4\cite{Schofield2022} & 50-200\cite{Schofield2022} & & 55\cite{Schofield2022}\footnotemark[1]\\
        DBT:DBN & co-sublimation & 755\cite{Moradi2019} & & 4.8\cite{Moradi2019} & 30-45\cite{Moradi2019} & &\\
        Tr:\textit{p}T\footnotemark[2] & spin-coating & 590\cite{Chu2017a} & & 4\cite{Harms1999} & & $>$95\cite{Buchler2005} &\\
        Tr:\textit{p}T & co-sublimation & 580\cite{Kummer1995} & & 4\cite{Kummer1995,Harms1999} & 40\cite{Kummer1995}, 50\cite{Hettich2002} & & \\
        DBATT:Shpol'skii\footnotemark[3] & shock-freezing & 590\cite{Boiron1996,Bloess2001} & 0.5\cite{Boiron1996},2\cite{Bloess2001} & 10\cite{Boiron1996,Rezai2018} & 20-25\cite{Boiron1996}, 12.5\cite{Kiefer2016,Rezai2018} & &\\
        DBATT:naphthalene & melting & 620\cite{Jelezko1997} & 0.5\cite{Jelezko1997} & 7\cite{Trebbia2022} & 15-25\cite{Jelezko1997}, 20-30\cite{Trebbia2022} & $\sim$100\cite{Trebbia2022} & 30\cite{Trebbia2022}\\
    \end{tabular}
    \end{ruledtabular}
    \footnotetext[1]{Not calibrated}
    \footnotetext[2]{Room temperature}
    \footnotetext[3]{\textit{n}-tetradecane or \textit{n}-hexadecane}
\end{table*}

\begin{figure*}
\includegraphics[width=16.5cm]{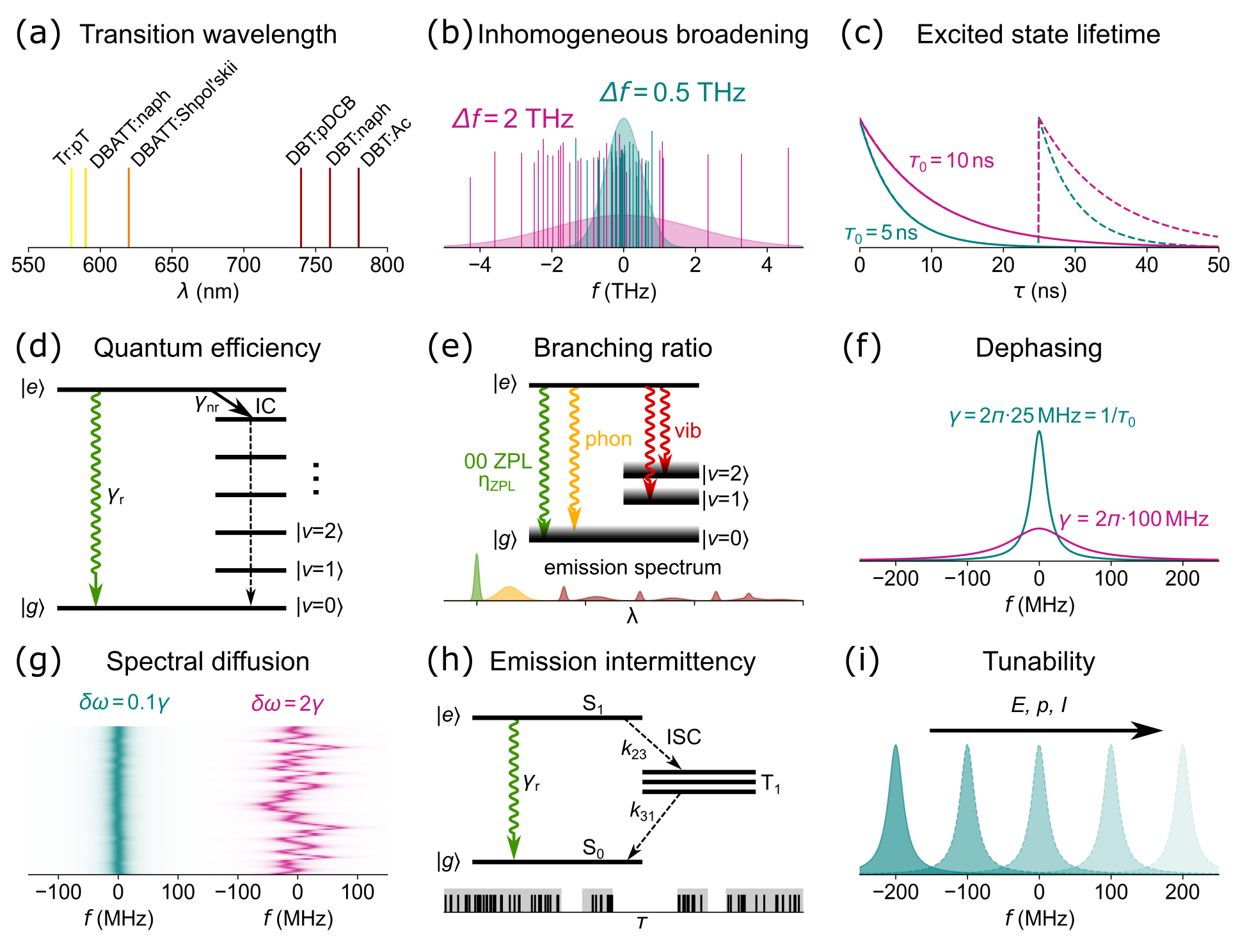}
\caption{\label{fig:properties_illustrations} Relevant emitter parameters. (a) Typical 00ZPL wavelengths spanning from \SI{580}{nm} to \SI{780}{nm} for different molecules and host matrices. (b) Simulated distribution of molecule frequencies for two different IHB widths $\Delta f$. (c) Simulated photon emission rate as a function of time for a pulsed photon source; solid lines illustrate \SI{50}{ns} repetition period, while dashed lines correspond to a \SI{25}{ns} period. (d) Jablonski diagram which includes internal conversion (IC), a non-radiative decay process with rate $\gamma_\mathrm{nr}$, which competes with the direct radiative decay at a rate $\gamma_\mathrm{r}$. (e) Jablonski diagram of different fluorescence emission processes: 00ZPL (resonant emission) in green, phonon wing in yellow, vibronic transitions in red. Below is an illustration of a corresponding emission spectrum. (f) Simulation of a single molecule excitation spectrum for two different homogeneous linewidths. The broader linewidth, corresponding to extra dephasing, produces distinguishable photons and requires higher excitation powers to reach the same emission level. (g) Simulations of repeated scans over two emitters with different magnitudes of spectral diffusion resulting in different amounts of sweep-to-sweep resonance frequency variation. (h) Simplified Jablonski diagram illustrating the transition to the long-lived triplet state $T_1$ via intersystem crossing (ISC). Here $k_{23}$ and $k_{31}$ are average rates into and out of the triplet state, respectively. Below is a simulated photon detection trace, where shaded time intervals correspond to a ``bright'' emitter in $S_0$-$S_1$ states, blank intervals denote a ``dark'' emitter in $T_1$ state, and black lines signify individual photon emissions. (i) Illustration of a single molecule resonance shifting under a controlled perturbation, e.g., an electric field (Stark tuning), pressure (strain tuning), or prolonged laser exposure (optical tuning).}
\end{figure*}

\subsection{Wavelength} Wavelength is a key parameter for an SPS, as it dictates material systems and optical transmission methods with which the source can effectively interface. The wavelength of molecular emitters depends both on the guest species and on the host matrix, and can be varied in a relatively large range by choosing a suitable guest-host combination. Commonly used organic molecules cover the longer wavelength part of the visible spectrum, ranging from \SI{580}{nm} for Tr in \textit{p}T\cite{Kummer1995} to \SI{780}{nm} for DBT in Ac\cite{Nicolet2007} (see Fig. \ref{fig:properties_illustrations}a and Table \ref{tab:material_systems}).

In general, longer wavelengths are preferred, as the required excitation light produces less fluorescence from surrounding materials, thus reducing the fluorescence background and improving the single-photon purity. This is especially true for room temperature applications, where single-photon emission is broadband. On the other hand, longer wavelength emitters generally have lower quantum efficiency\cite{Englman1970,Erker2022} (see subsection \ref{subsec:emitter_parameters_QE}). This sets a practical limit to about \SI{800}{nm}. Finally, some specific wavelength can be of particular interest for interfacing with other quantum systems. For example, the 00ZPL of DBATT in \textit{n}-tetradecane at \SI{589}{nm} matches the sodium D-line transition\cite{Siyushev2014,Kiefer2016}; in addition to being potentially useful as a quantum memory, it can also serve as a very narrow-band spectral filter for the 00ZPL emission, improving the photon indistinguishability\cite{Kiefer2016,Rezai2018}.

\subsection{Inhomogeneous broadening} As mentioned in section \ref{sec:overview}, the wavelength of individual molecules in the same host material can vary significantly one from the other (see Fig. \ref{fig:overview}f and Fig. \ref{fig:properties_illustrations}b) due to IHB. This is especially apparent in low-temperature applications, where this variation can in extreme cases be 5-6 orders of magnitude larger than the transition linewidth of the molecules. This spread arises from variations of the local environment, such as pressure\cite{Muller1995,Gmeiner2016} or local electric fields\cite{Wild1992,Orrit1992} which shift the resonance frequencies of different molecules to a different degree. The IHB width thus depends on the host material, with highly ordered crystals displaying relatively narrow IHB of several tens or hundreds of GHz\cite{Nicolet2007,Ambrose1991}, while highly-disordered polymer matrices produce IHB of up to about \SI{10}{THz}\cite{Orrit1992,Walser2009,Rattenbacher2023}. Even for the same guest-host combination, the IHB width can vary depending on the sample preparation methods and even on the details such as crystal growth speed; e.g., the same DBT:Ac system has an IHB width of \SI{150}{GHz} for high-quality co-sublimated crystals\cite{Nicolet2007} and about \SI{10}{THz} for nanocrystals\cite{Musavinezhad2024}.

The optimal width often depends on the application: larger transition wavelength spread allows for easier addressing of individual emitters even in relatively dense samples, simplifying selection of optimally positioned and oriented molecules. Furthermore, it gives more flexibility in selecting emitters at a specific frequency, e.g., for interfacing with a different optical system. On the other hand, if several emitters should be at precisely the same frequency (e.g., for simultaneous indistinguishable photon emission\cite{Lettow2010}), narrower IHB width presents an advantage.

\subsection{Excited state lifetime} The excited state lifetime $\tau_0$ (also often referred to as $T_1$) defines the time the emitter spends on average in the excited state $\ket{e}$ before decaying back to the ground state $\ket{g}$, and it effectively defines the ``reset speed'' of the emitter. In a CW single-photon source it determines the maximal photon emission rate, which is equal to $1/(2\tau_0)$ for resonant drive and $1/\tau_0$ for higher-energy excitation, provided the quantum efficiency is close to unity. In the case of a pulsed source, it ultimately limits the maximal pulse repetition rate (see Fig. \ref{fig:properties_illustrations}c).

As molecular emitters have a high quantum efficiency, their lifetime is mainly determined by the radiative decay rate, which is ultimately a function of the transition dipole moment and the refractive index of the surrounding host material. Both of these have relatively low variations between different molecule and host species: the former is mainly determined by the geometrical properties of the molecule and its orbitals, and the latter is confined in a relatively narrow range between 1.5 and 2 for most organic hosts. As a consequence, the lifetime shows little variations, changing from \SI{4}{ns} for DBT in Ac\cite{Pazzagli2018,Ren2022} or \textit{p}T\cite{Schofield2022} and Tr in \textit{p}T\cite{Harms1999,Kummer1995} to \SI{10}{ns} for DBATT in \textit{n}-hexadecane\cite{Boiron1996}. It also stays fairly constant between room and cryogenic temperatures\cite{Harms1999}, further confirming mostly radiative origin.

\subsection{Quantum efficiency}
\label{subsec:emitter_parameters_QE}
When an emitter is in its excited state, it normally decays radiatively, i.e., via emission of a photon. However, other, non-radiative, processes are also possible. Their contribution is usually characterized via the quantum efficiency $\eta=k_\mathrm{r}/(k_\mathrm{nr}+k_\mathrm{r})$, where $k_\mathrm{r}$ and $k_\mathrm{nr}$ are the radiative and non-radiative decay rates, respectively (see Fig. \ref{fig:properties_illustrations}d). The excited state lifetime is then determined by the combined decay rate and is expressed as $\tau_0=1/(k_\mathrm{nr}+k_\mathrm{r})$.

These non-radiative decay processes are especially detrimental to pulsed SPS, as they fundamentally limit the probability of generating a single photon upon emitter excitation to be at most $\eta$. Furthermore, they reduce the excitation efficiency of the emitter, which necessitates higher excitation powers and, consequently, higher background fluorescence levels.
CW single-photon sources are less affected by additional non-radiative decay channels, as the maximal photon emission rate is still determined purely by the radiative decay rate; nevertheless, the higher excitation power requirement applies there as well.

In PAH molecules the main currently identified non-radiative decay rate is internal conversion (IC), which is the direct conversion of the electronic excitation into several intramolecular vibrations without an associated photon emission (see Fig. \ref{fig:properties_illustrations}d). The efficiency of this process has been found to obey an energy gap law\cite{Englman1970,Erker2022}, which states that lower energy excitations generally have higher IC rates: $k_\mathrm{IC}\propto \exp(-\beta \Delta E)$, where $k_\mathrm{IC}$ is the IC rate, $\Delta E=\hbar\omega_{eg}$ is the energy of the excited state at the transition frequency $\omega_{eg}$, and $\beta$ is a proportionality constant depending on the molecule's vibrational properties. Thus, this process is mostly relevant to longer-wavelength emitters such as DBT. Indeed, it has been shown that the quantum efficiency of Tr\cite{Buchler2005, Chu2017a} and DBATT\cite{Trebbia2022} (both at wavelengths around \SI{600}{nm}) can be very close to 1, while for DBT (typical wavelength between \SI{740}{nm} and \SI{780}{nm}) the room temperature estimates lie around 0.35\cite{Erker2022}, while cryogenic measurements indicate $\eta\geq0.7$\cite{Musavinezhad2023}. The discrepancy between the two measurements is not completely understood at present, but it has been theoretically predicted, that the non-radiative decay rates might decrease at cryogenic temperatures\cite{Bassler2024}.

Since the energy of the molecular vibrations is much lower than the electronic excitation energy, the IC is necessarily a higher order process involving the creation of several vibrational excitations. Therefore, it typically involves the highest frequency modes, which in PAH correspond to C-H bond stretching at a frequency of about \SI{93}{THz}, about 1/4 of the electronic transition frequency. This suggests a way to reduce the IC efficiency by deuteration, whereby the hydrogen (protium) atoms along the perimeter of the molecule are substituted for heavier deuterium atoms. This replaces C-H bonds with C-D, whose frequency is significantly (about $\sqrt{2}$, as the D mass is twice that of H) lower at \SI{66}{THz}, increasing the required number of vibrational excitations, making the IC process less likely. Indeed, it has been demonstrated that deuteration can reduce the non-radiative decay rate by about a factor of two for DBT \cite{Mishra2024}, demonstrating a way to achieve unity quantum efficiency.

\subsection{Branching ratio}
In addition to the difference between radiative and non-radiative processes, one often needs to distinguish between the resonant 00ZPL emission and the red-shifted fluorescence associated with the excitation of matrix phonons or molecular vibrations (see Fig. \ref{fig:properties_illustrations}e and discussion in section \ref{sec:overview}). As these red-shifted photons are spread over a very large spectral range, they are not indistinguishable, so they must be filtered out to preserve the photon indistinguishability. Since these photons are removed, the photon flux is reduced; in this sense, the 00ZPL branching ratio adds an additional factor to the quantum efficiency when indistinguishable photons are desired. One should note, however, that not all applications require this filtering, and in some cases it is sufficient to have distinguishable photons of somewhat larger wavelength spread\cite{Lombardi2020}.

The fraction of the 00ZPL in the total emission is called the branching ratio $\alpha_\mathrm{ZPL}$; it is often further split into Franck-Condon and Debye-Waller factors describing the effects of the molecular vibrations and matrix phonons respectively. The Franck-Condon factors $F_{i,n}$ are normally defined for individual intramolecular vibrational modes, and they show the probability of creating $n$ vibrational quanta in the mode $i$ upon decaying from the vibrational ground state of the electronically excited state to the electronic ground state. They are calculated as an overlap integral between the vibrational wavefunctions in the excited and ground electronics states: $F_{i,n}=\left|\langle g;v_i=n|e;v_i=0\rangle\right|^2$, where $\ket{s;v_i=n}$ corresponds to a state with $n$ quanta in the vibrational mode $i$ in the electronic state $s$ (ground or excited). In the simplest case, the vibrational modes have the same frequencies and shapes in both ground and excited states, so the overlap integral arises purely due to a displacement between the equilibrium positions of the mechanical mode in the two electronic states. In other words, the equilibrium molecule configuration in the electronic excited state corresponds to a shifted configuration in the electronic ground state, i.e., coherent vibrational state. Most often one is interested in the 00ZPL Franck-Condon factor, which can be found by combining the contributions from all of the individual vibrational modes: $F_\mathrm{0}=\prod_i F_{i,0}=e^{-\sum_i \Lambda_i^2}$.

The precise experimental determination of the Franck-Condon factor is technologically challenging, as it requires intensity calibrated measurements of emission over a wavelength range of \SI{200}{nm} (frequency range of \SI{100}{THz}) coming from a cryogenic sample, typically a single molecule or a small ensemble of molecules. Therefore, such measurements are relatively sparse in the literature. Generally, the higher reported values such as \SI{44}{\percent} for DBT in \textit{p}DCB\cite{Verhart2016}, \SI{55}{\percent} for DBT in \textit{p}T\cite{Schofield2022}, and \SI{46}{\percent} for DBT in Ac, do not correct for chromatic aberrations in the setup or only observe a part of the spectrum, likely overestimating the 00ZPL branching ratio; more calibrated measurements typically produce somewhat lower results, such as \SI{33}{\percent} for DBT in Ac\cite{Trebbia2009} and \SI{30}{\percent} for DBATT in naphthalene\cite{Trebbia2022}. Further complicating the matter, a recent study for Tr on an hBN surface\cite{DeHaas2025} demonstrated a large variation of the branching ratio for different individual emitters, which was attributed to nearby defects in hBN modifying the charge distribution in Tr molecules and, consequently, its Franck-Condon factor. This suggests potentially high variability of the 00ZPL branching ratio between individual emitters in the same sample, although the conservative lower bound of \SI{30}{\percent} has been reliably demonstrated for all of the observed systems.

\subsection{Dephasing}
\label{subsec:emitter_parameters_dephasing}
Just as the frequencies of individual emitters are different due to spatial inhomogeneity of the host material, so can the individual emitter central frequency change in time due to fluctuations of the environment. When these fluctuations happen faster than the excited state lifetime, they result in dephasing, that is, reduced coherence of the emitter and, correspondingly, of the emitted photons. As a consequence, the photon indistinguishability is reduced\cite{Rezai2018,Lombardi2021,Schofield2022b}. The effects of dephasing are typically identified and characterized by comparing the spectroscopic linewidth (full width at half-maximum) of the emitter $\gamma$ with the so-called lifetime limit (also often referred to as Fourier limit or radiative limit) $\gamma_0=1/\tau_0$, which is the narrowest linewidth possible in the absence of dephasing (see Fig. \ref{fig:properties_illustrations}f). If $\gamma\approx\gamma_0$, the emitter is considered to be lifetime-limited. In different contexts one often uses the photon coherence time $T_2=2/\gamma$ instead of spectroscopic linewidth $\gamma$; in this case, the lifetime limit $\gamma=\gamma_0$ corresponds to the well-known relation $T_2=2T_1$.

Dephasing in the case of molecules originates from two sources. The first is the thermally fluctuating environment, typically described in terms of phonon-induced dephasing. This leads to temperature-induced broadening\cite{Kummer1995b,Plakhotnik1997}, which is fairly well understood\cite{Skinner1988,Clear2020,Reitz2020}, to the point where it has been recently used for molecule-based temperature sensing\cite{Esteso2023}. This mechanism is the main reason why cryogenic temperatures are a necessary requirement to produce indistinguishable photons. Molecular emitters typically require temperatures between \SI{2}{K}\cite{Kummer1995b} and \SI{4}{K}\cite{Jelezko1997,Nicolet2007} in order to reach lifetime-limited linewidths, where the exact value depends on the guest molecule, the host material, and on the specific insertion site of the guest within the host\cite{Nicolet2007}.

The second dephasing source comes from other low-energy excitations of the host material, which are generally much less explored and understood. This manifests as residual spectral broadening present even at the temperatures where phonon-induced dephasing is expected to be negligible, and is often attributed to localized crystal defects or surface imperfections. As a general rule, one observes stronger dephasing in less stable and more disordered host materials. For example, molecules at surfaces\cite{Smit2022} or in polymers such as PMMA\cite{Walser2009} or PE\cite{Donley2000a,Rattenbacher2023} can display linewidths an order of magnitude above the lifetime limit, while molecules in well-ordered organic crystals such as Ac\cite{Nicolet2007} or \textit{p}DCB\cite{Verhart2016,Musavinezhad2023} routinely reach the lifetime limit at \SI{2}{K}. Nanocrystals, due to their small size and potentially large inhomogeneity, lie somewhere in between, with molecules demonstrating linewidths typically about twice the lifetime limit\cite{Pazzagli2018,Musavinezhad2024}. It is worth noting that some of this extra dephasing can still be reduced by lowering the temperature to the milli-Kelvin regime, as has been demonstrated for PE with Tr\cite{Donley2000a} and DBT\cite{Rattenbacher2023}.

Overall, all the cryogenic molecular material systems considered for SPS application are operated in the regime where the dephasing is significantly lower than the linewidth or is completely absent, as routinely demonstrated through high photon indistinguishability\cite{Rezai2018} and direct measurement of Rabi oscillations in time domain\cite{Gerhardt2009,Rezai2019c,Deplano2023}. The latter has been used to demonstrate an excitation probability of up to \SI{97}{\percent} using a coherent $\pi$-pulse (i.e., driving half of a Rabi oscillation period)\cite{Deplano2023}, which shows its potential for replacing 0-1 excitation for deterministic single-photon generation.

\subsection{Spectral diffusion and stability}
When the characteristic time of frequency fluctuations is slower than the excited state lifetime, one usually refers to them as spectral diffusion (See Fig. \ref{fig:properties_illustrations}g). The limitations that this diffusion places on the single-photon emitter properties are more complex. On one hand, if this diffusion is slow compared to the emission rate, it does not reduce indistinguishability of consecutive photons, as they still have very similar frequency. On the other hand, if the emitter is excited via its narrow 00ZPL, the spectral diffusion can lead to fluctuations of excitation efficiency, which can show up as intensity variations in CW drive, or excitation pulse area errors in the case of pulsed resonant excitation. In addition, it can still reduce indistinguishability between photons emitted by separate molecules, as it causes constant changes between their resonance frequencies.

The general considerations here are similar to the case of dephasing. Less ordered systems such as nanocrystals\cite{Pazzagli2018} or polymers\cite{Rattenbacher2023} frequently result in less stable emission frequencies. The observed spectral diffusion is often at least partially photo-activated, i.e., it comes about when the whole crystal or the emitter itself are excited by light. These effects pose problems for SPS applications, where one typically strives to excite the emitter as strongly and consistently as possible. On the other hand, in higher quality crystalline materials such as Ac\cite{Ren2022} and \textit{p}DCB\cite{Zirkelbach2022} spectral diffusion is negligible, with spectral fluctuations of much less than a linewidth on the timescale of minutes or hours under strong continuous illumination.

\subsection{Emission intermittency} 
In addition to the frequency instability, some emitters also exhibit fluctuations of their intensity. These most often manifest as blinking, which is a temporary reduction of emission, as opposed to bleaching, which is a permanent loss of optical activity. There are two main causes of blinking in molecular quantum emitters at low temperatures. The first is an apparent blinking arising upon resonant 00ZPL excitation of an emitter experiencing large reversible spectral jumps; if the emitter's transition frequency changes by much more than its excitation linewidth, it stops being efficiently excited, effectively going dark until the frequency is restored. The other commonly observed cause of blinking is the transition from the lowest electronically excited singlet state $S_1$ into a longer-lived triplet state $T_1$ through a process referred to as intersystem crossing (ISC; see Fig. \ref{fig:properties_illustrations}h). While in the triplet state, the emitter is no longer optically active, resulting in a temporary loss of emission until the triplet state decays back into the singlet ground state.

ISC is characterized by two parameters: the rate of the crossing from the excited singlet to the triplet state $k_{23}$ (often specified by the triplet yield $k_{23}\tau_0$, which is the probability of transitioning into the triplet state upon a single emitter excitation) and the triplet state decay rate $k_{31}$ (see Fig. \ref{fig:properties_illustrations}h). The effect of ISC on the emission depends on the excitation strength. In the strong excitation limit the triplet state occupation probability is about $k_{23}/(k_{31}+k_{23})$, with the corresponding drop in the photon emission rate. In the opposite weak limit, where the emission rate is much lower than $k_{23}$, the triplet state has enough time to decay between the two consecutive excitation events, resulting in the negligible ISC effect, on the order of the triplet yield $k_{23}\tau_0\ll k_{23}/(k_{31}+k_{23})$.

Historically, the first molecular emitter system (pentacene in \textit{p}T\cite{Moerner1989}) had relatively high triplet state yield in the range of \SI{0.5}{\percent} to \SI{60}{\percent}, strongly depending on the crystal site. Combined with a long triplet state lifetime of around \SI{10}{\micro s} to \SI{100}{\micro s}\cite{Kohler1996,Brouwer1999}, this resulted in a significantly reduced emission rate. In contrast, most currently used dyes, such as DBT, Tr and DBATT, have triplet yields around $10^{-5}$ to $10^{-7}$ and triplet state lifetimes between \SI{10}{\micro s} and \SI{10}{ms}, with the longer-lived state typically having lower yield.\cite{Nicolet2007,Musavinezhad2023,Kummer1995,Boiron1996,Jelezko1997}. This results in an emission reduction between \SI{0.1}{\percent} and \SI{1}{\percent} even at the maximal excitation rate, which can be neglected for most applications.

\subsection{Tunability}
In some applications it is convenient to be able to control the frequency of the single photon emission. As discussed above, coarse ``tuning'' can be achieved by selecting the appropriate guest-host system and addressing a particular emitter within the IHB. However, often finer tuning is needed, e.g., to match to some external narrow-band system such as a fixed cavity or a different emitter\cite{Rezus2012} (see Fig. \ref{fig:properties_illustrations}i), or when one aims to generate indistinguishable photons from several independent emitters\cite{Lettow2010,Duquennoy2022}.

The most common way to achieve this with molecular emitters is though the use of Stark tuning\cite{Wild1992,Orrit1992}, which is the shift of the emission frequency upon application of an electric field. This method is very versatile, with the source of electric field ranging from simple micro-fabricated electrodes\cite{Bauer2002,Lettow2010,Rattenbacher2024,Trebbia2022,Huang2025} to 2D materials\cite{Schadler2019} or micron-scale tips\cite{Hettich2002}. Thanks to its simplicity and compatibility with on-chip photonic structures, it has been used in several experiments for tuning two emitters in order to generate indistinguishable photons\cite{Lettow2010,Huang2025}. Since most common PAHs (DBT, DBATT, Tr, pentacene) have inversion symmetry, their linear Stark coefficient is nominally zero, so the most pronounced effect comes from the quadratic shift. However, in practice, these molecules very often display a non-zero linear Stark coefficient\cite{Orrit1992,Brunel1999}, which is usually attributed to the inversion symmetry breaking due to deformations of the molecule and strong static electric fields present in the host material. As this effect is different from one molecule to another, it allows for efficient tuning of separate molecules using a single control voltage \cite{Trebbia2022,Huang2025}. Furthermore, specific host matrices such as dibromonaphthalene (DBN), whose constituent molecules and crystal structure lack inversion symmetry, can induce a persistent strong linear dipole moment in the guest molecules\cite{Moradi2019}. This greatly improves the linear tuning coefficient and, as a consequence, the maximal tuning range.

A different, recently discovered tuning method uses resonant or off-resonant laser excitation to semi-permanently shift emitters' resonance frequencies\cite{Colautti2020,Duquennoy2022,Lange2024,Nobakht2025}. The mechanism is not definitively identified at the moment, but is thought to be related to a local matrix polarization caused by the energy delivered to the matrix via the laser excitation. This shift typically persists for hours or days and has a magnitude in the order of \SI{100}{GHz}, which is much larger than normally achieved with simple electrode Stark tuning. As an additional advantage, it does not require electrode fabrication, so it can be applied to any molecule within the sample. On the other hand, it is harder to precisely control the frequency shift with this approach, and it cannot be easily applied dynamically. Nevertheless, its simplicity and versatility have already allowed for bringing two molecules in resonance to implement near-field coupling\cite{Lange2024} or the generation of indistinguishable photons\cite{Duquennoy2022}. Recently, the combination of both optical tuning and Stark tuning methods\cite{Duquennoy2024} has been demonstrated to provide a finer control of the local electrical environment of a molecule to simultaneously tune its frequency and minimize spectral diffusion.

Finally, one can also realize tuning using strain, which is also known to affect the resonance frequency\cite{Croci1993,Muller1995,Tian2014} at a level of about \SI{1}{THz} for strain of \SI{1}{\percent}. Currently this method is still being developed, and the demonstrated frequency shifts are somewhat lower than those achieved via the Stark effect\cite{Fasoulakis2022}. Nevertheless, it potentially promises significant tunability of up to several 100's of \SI{}{GHz} for optimized piezoelectric materials and strain geometries\cite{Fasoulakis2022}.

\section{Photon collection methods}
\label{sec:collection_methods}

\begin{figure*}
\includegraphics[width=16.5cm]{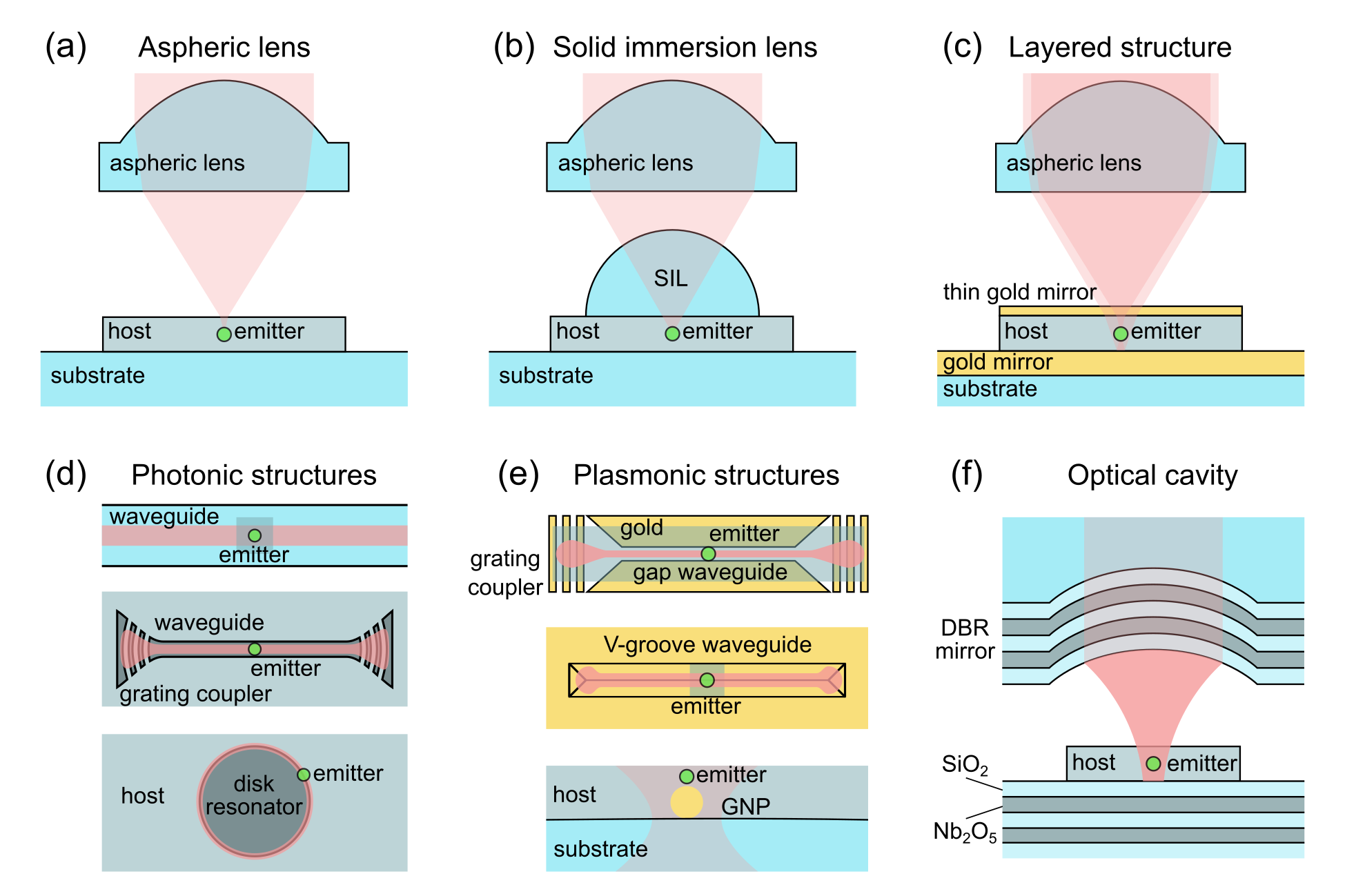}
\caption{\label{fig:extraction_illustrations} Most common photon collection methods used with molecular single photon emitters. (a) Aspheric lens or cryogenic objective directly collecting emission from a molecule embedded into a host material. (b) Solid immersion lens (SIL) used to increase the collection angle \cite{Wrigge2008,Trebbia2009,Siyushev2014,Rezai2018,Colautti2020b,Zirkelbach2022}. (c) Additional layers can be used for engineering planar antenna structures\cite{Lee2011,Checcucci2016,Chu2017a,Wei2020,Lombardi2020,Murtaza2023}. (d) Free-standing optical waveguides (top) or on-chip dielectric photonic structures such as sub-wavelength waveguides (middle) or resonators (bottom) used to route emission\cite{Turschmann2017,Lombardi2018,Rattenbacher2019,Boissier2021,Shkarin2021,Colautti2020b,Ren2022,Rattenbacher2023}. (e) Plasmonic gap (top) or v-groove (middle) waveguides and plasmonic gold nanoparticles fabricated on a glass surface (bottom)\cite{Grandi2019,Kumar2020,Zirkelbach2020}. (f) Fabry-Perot cavity funneling emission into the cavity mode\cite{Wang2017,Wang2019}.
}
\end{figure*}

As described above, there are several organic material platforms demonstrating properties which are well-suited for SPS applications. An outstanding challenge is efficient photon collection. There, one aims to achieve the highest possible collection efficiency $\eta_\mathrm{coll}$, defined as the ratio of collected photons to total emitted photons.

\subsection{Free space optics}

At room temperature, high photon collection efficiency is straightforward due to the availability of high numerical aperture (NA) oil immersion objectives. Commercial coverglasses in combination with such an objective allow for photon collection efficiencies on the order of 80\%, if the molecule is located in a film of thickness below several 10's of nanometers. One exploits here the fact that the interface shifts the radiation pattern of the molecule towards a higher refractive index material, i.e., the cover glass and, thus, the microscope objective \cite{Novotny2012}.

In cryogenic experiments, the situation is more challenging due to the lack of high NA objectives. With an aspheric lens with an NA of 0.77 (see Fig. \ref{fig:extraction_illustrations}a) the collection efficiency is about 1 to 10\%, strongly depending on the dipole orientation. The situation can be dramatically improved by combining the objective with a so-called solid immersion lens (SIL)\cite{Lettow2010,Siyushev2014,Colautti2020b}, which is nothing but a half-ball lens with high refractive index (see Fig. \ref{fig:extraction_illustrations}b). Typically, these lenses are made of cubic zirconia with a refractive index of 2.15, but other materials such as diamond or gallium phosphide have been used as well\cite{Lettow2010,Yurgens2021,Chen2018}. The advantage of the SIL is twofold. First, the radiation pattern of a molecule close to the SIL surface will again be strongly modified such that the emission is mainly directed towards the collection optics. Second, all photons collected by the SIL will impinge by design on the curved spherical surface at an angle of zero degrees and will thus not be refracted as the photons leave the half sphere. As a consequence, images are not distorted and total internal reflection is avoided. Furthermore, this can be enhanced with an anti-reflection coating to avoid large Fresnel reflection losses due to the high refractive index of the SIL. With the same collection lens with an NA of 0.77 and a refractive index of the host material of 1.6, a SIL with $n=2.15$ can improve the collection efficiency from 1-10\% to almost 50\%. Photon detection rates on the order of a few millions per second have been reported in such a system\cite{Maser2017}. Note, that a so-called Weierstrass lenses could, in principle, improve the collection efficiency further \cite{Barnes2002}.

Further improvements, up to near-unity collection efficiency, can be achieved with a planar antenna (see Fig. \ref{fig:extraction_illustrations}c), which consists in its simplest realization out of a few hundred nanometer thick film containing the emitters placed on top of a high-index substrate. This layered structure modifies the radiation pattern in such a way that photons are preferentially emitted into the high-index substrate under an angle limited by Snell's law \cite{Lee2011}. These antennas allow for unprecedented photon collection and detection rates \cite{Lee2011,Chen2011,Chu2014,Checcucci2016}. End-to-end efficiencies on the order of $70$\% lead to highly efficient single-photon sources with a measured intensity squeezing as high as \SI{2.2}{dB}\cite{Chu2017a}, i.e., with the intensity fluctuations reduced \SI{40}{\percent} below the shot noise limit owing to the regular arrival times of individual photons. These types of devices have been further developed into truncated metallo-dielectric omnidirectional reflectors, such that efficient emission into a single-mode fiber with about 95\% efficiency is predicted \cite{Li2020}. Planar antennas are very versatile and can be implemented in various environments, including liquids \cite{Morales2024}.

\subsection{Integrated photonics and plasmonics}

While free-space methods boast extremely high collection efficiency, they often come with complicated spatial mode profiles, which are challenging to couple to single-mode fibers or chip-based structures. While this can, potentially, be addressed by using spatial light modulators (SLMs) or metasurfaces\cite{Li2023}, neither of these methods is ideal: SLMs often have relatively high loss of 30-\SI{50}{\percent}, while metasurfaces lack in-situ tunability and must be tailored to a particular sample and, possibly, to a particular emitter. Furthermore, the highest collection efficiency has been demonstrated at room temperature, and a cryogenic implementation of similar efficiency is still lacking.

One way to solve the spatial profile issue is photonic integration, where one aims to efficiently capture single photons into free-standing waveguides or on-chip photonic structures (see Fig. \ref{fig:extraction_illustrations}d). This allows them to either be directly integrated into photonic circuits or coupled to optical fibers with high efficiency, leveraging existing off-chip coupling strategies\cite{Son2018}. Here, the main challenge is the current lack of host materials which are directly compatible with photonic structure fabrication methods. The most common molecule host materials are molecular crystals and Shpol'skii matrices, which cannot be easily shaped with nanometer precision, and which are either liquid or volatile at room temperature, evaporating sometimes at rates of \SI{>1}{\micro m/hr}. Alternative materials such as polymers are more stable, and some of them, such as PMMA, can be shaped using modern lithographic techniques. On the other hand, they have a relatively low refractive index comparable to that of the substrate, and they come at a cost of significantly degraded emitter quality. This necessitates hybrid techniques, where the structures are fabricated from a high refractive index material and the molecule-doped host surrounds them. The emitter-waveguide coupling in such structures is, in most cases, evanescent. 

This hybrid approach usually leverages high molecule density to implement a ``stochastic integration'': while the positioning of each individual molecule can not be controlled with nanometer precision, waveguides typically have hundreds or thousands of coupled molecules (corresponding to the linear density of tens to hundreds per micron), ensuring that many of them will be close enough to the surface to demonstrate high collection efficiency. While the evanescent coupling makes it more challenging to optimize the photon collection rate compared to integrating emitters directly into the photonic structures, this approach provides a flexibility advantage. The refractive index of the host material is relatively low, around 1.5 for many materials and axes in birefringent hosts, and below 2 in almost all cases. This simplifies light extraction and allows for direct integration with a variety of different dielectrics, simplifying the hybrid system design.

Several works have demonstrated structures with waveguides made of $\mathrm{TiO_2}$\cite{Turschmann2017}, $\mathrm{Si_3N_4}$\cite{Lombardi2018,Boissier2021,Ren2022} or GaP\cite{Shkarin2021}. The maximal collection efficiency was demonstrated to be about \SI{40}{\percent} at room temperature based on the observed emission rate\cite{Lombardi2018}, or about \SI{20}{\percent} at cryogenic temperatures based on extinction measurements\cite{Turschmann2017,Shkarin2021}, with simulations also suggesting up to \SI{40}{\percent} under optimal conditions\cite{Turschmann2017,Boissier2021,Shkarin2021,Ren2022}. The source of this discrepancy is unclear at the moment, but it is tentatively attributed to the minimal achievable distance from the emitter to the photonic structure surface, which suggests that further improvement is possible. One additional advantage of the waveguide-based collection technique is that it allows for efficient spatial or directional separation of the emission from the excitation light whenever spectral filters cannot be used, e.g., when the excitation is done using coherent $\pi$-pulses. This approach has been recently used to demonstrate a signal-to-background ratio of more than 100 in combination with polarization filtering\cite{Ren2022,Huang2025}.

Some works have also exploited the plasmonic emission enhancement by coupling single molecules to plasmonic waveguides at room temperature\cite{Grandi2019,Kumar2020} (see Fig. \ref{fig:extraction_illustrations}e). These have demonstrated a noticeable emission rate enhancement and comparable bi-directional coupling efficiency estimated to be up to \SI{50}{\percent}\cite{Kumar2020}; however, such waveguides still suffer from relatively short plasmon propagation length on the order of \SI{10}{\micro m}, hindering their incorporation into more complicated photonic circuits. As an alternative, hybrid plasmonic-dielectric structures have been proposed and realized\cite{Kewes2016}, but not yet tested with emitters.
Finally, plasmonic antennas have been employed at cryogenic temperatures to enhance the emitter decay rate, increasing it five-fold and producing photon detection rates of up to \SI{8}{Mcts/s} with the addition of a SIL\cite{Zirkelbach2020}.

\subsection{Purcell enhancement}

Light extraction can be further boosted by resonant enhancement of emission into a particular spatial mode, which can be engineered to allow for easier interfacing with other systems. For example, if a Fabry-Perot resonator is used, its output mode is a Gaussian, allowing for straightforward coupling to single-mode fibers. Alternatively, on-chip resonator structures can guide their emission directly into on-chip waveguides, providing a direct way to interface with other integrated photonic structures.

In addition to spatial mode shaping, the resonant structures can also affect the emission spectrum of the molecular emitter due to Purcell enhancement. Given a high enough resonant quality factor, one can engineer the resonator to increase only the 00ZPL emission rate, which can drastically diminish negative effects like non-unity quantum efficiency and 00ZPL branching ratio, turning the molecule into an ideal two-level emitter. The main figure of merit here is the Purcell factor $F$ describing the enhancement of the radiative 00ZPL transition. Given the initial 00ZPL efficiency $\eta_\mathrm{ZPL}=\eta\alpha_\mathrm{ZPL}$ (which includes both the QE and the branching ratio), the total decay rate can be decomposed as a sum $\gamma_0=\gamma_\mathrm{ZPL}+\gamma_\mathrm{loss}$ of the 00ZPL radiative rate of $\gamma_\mathrm{ZPL}=\eta_\mathrm{ZPL}\gamma_0$ and the remaining decay processes $\gamma_\mathrm{loss}=(1-\eta_\mathrm{ZPL})\gamma_0$. As the Purcell enhancement factor only applies to the 00ZPL, the resulting enhanced decay rate is $\gamma_0^*=F\gamma_\mathrm{ZPL}+\gamma_\mathrm{loss}=(1+(F-1)\eta_\mathrm{ZPL})\gamma_0$, and the enhanced 00ZPL efficiency is $\eta_\mathrm{ZPL}^*=\eta_\mathrm{ZPL}F/(1+(F-1)\eta_\mathrm{ZPL})$, which approaches 1 for $F\eta_\mathrm{ZPL}\gg1$. As an additional benefit, the enhanced lifetime-limited decay rate $\gamma_0^*$ reduces the effects of a potential residual dephasing, allowing operation at higher temperatures and in a larger range of host materials.

Such enhancement has been decisively demonstrated using an open Fabry-Perot microcavity\cite{Wang2019,Pscherer2021} (see Fig. \ref{fig:extraction_illustrations}f), where the achieved $F\approx 40$ enhanced the 00ZPL efficiency from \SI{30}{\percent} to \SI{95}{\percent}, simultaneously increasing the total decay rate by about a factor of 15 and entering the strong coupling regime of cavity QED. In addition to bringing the molecules' properties close to the ideal two-level emitter and, consequently, ideal single-photon source, the cavity coupling can also be used in its own right to explore other non-classical phenomena in light-matter interaction. As an example, the Fabry-Perot microcavity system has already been used for the demonstration of non-linear optical effects such as four-wave mixing and optical switching at a single-photon level\cite{Pscherer2021}.

The other advantage of a cavity comes from the fact that the molecule's emission is collected through its mode, so the spatial mode matching becomes much more straightforward. However, this approach is significantly more technologically challenging due to the very stringent vibrational stability requirements: for a finesse of 20,000 in the work above, a change in the cavity length by \SI{20}{pm} detunes it by a linewidth, effectively decoupling the emitter. Even smaller displacements affect the detuning and, consequently, the Purcell enhancement, which in turn changes the emitter lifetime and reduces the photon indistinguishability. While a monolithic Fabry-Perot cavity design\cite{Press2007} can eliminate mechanical noise, the same degree of enhancement has not yet been demonstrated in such structures with molecular emitters. Furthermore, in such microscopic cavities, one generally can only optically access the emitter through the very spectrally selective cavity mode, which poses challenges with 0-1 excitation schemes. As a result, it requires specialized mirror coatings, which are highly reflective at the 00ZPL wavelength but have high transmission for the excitation light.

As an alternative, a chip-based approach provides higher stability and easier independent access to the emitters; however, it relies on an evanescent coupling, typically resulting in lower Purcell enhancement. On top of that, in many cases, the host crystal in direct contact with the photonic structure may lead to additional scattering losses, further reducing the Purcell enhancement\cite{Rattenbacher2019}. Two different approaches have been demonstrated recently to address these issues. In one, on-chip disc resonators were used in conjunction with a homogeneous and transparent PE host material, which resulted in a finesse of about 200. This allowed for a Purcell enhancement of $F\approx 10$, increasing the overall emitter decay rate by about a factor of three\cite{Rattenbacher2023}. However, the use of a polymer PE matrix resulted in a photoinduced spectral diffusion, currently preventing it from being a viable single-photon source. The other recently demonstrated approach relies on photonic crystal cavities coupled to a co-sublimated DBT:Ac crystal placed on top of it. While it demonstrates slightly lower Purcell enhancement of $F\approx 3.5$, the crystalline host material is much more suitable for the production of stable emitters, making it very promising for SPS applications.

\section{Performance of single-molecule single-photon sources}
\label{sec:current_state}

\begin{table*}
    \caption{\label{tab:figures_of_merit}Figures of merit for molecular single-photon sources}
    \begin{ruledtabular}
    \begin{tabular}{p{4cm}|p{3.5cm}|p{2cm}|p{5cm}|p{2cm}}
        Parameter & System & Value & Comment & References \\\hline\hline
        \multirow{4}{3cm}{Purity, $g^{(2)}(0)$} & DBT:Ac nanocrystals & $0.03\pm 0.01$ & RT\footnotemark[1], CW\footnotemark[1] & \onlinecite{Pazzagli2018} \\
         & DBT:Ac nanocrystals & $0.02$ & RT, pulsed\footnotemark[1] & \onlinecite{Murtaza2023} \\
         & DBATT:\textit{n}-tetradecane & $0.02_{-0.00}^{+0.03}$ & LT\footnotemark[1], CW, atom vapor filter & \onlinecite{Rezai2018} \\
         & DBT:Ac nanocrystals & $0.003\pm 0.01$ & LT, pulsed & \onlinecite{Duquennoy2022} \\\hline
         \multirow{2}{3cm}{Collection efficiency} & Tr:\textit{p}T & $>0.96$ & RT, CW, planar antenna & \onlinecite{Lee2011,Chu2017a}\\
         & DBATT:naph & $0.43$ & LT, CW, SIL & \onlinecite{Maser2017}
         \\\hline
         End-to-end efficiency & Tr:\textit{p}T & $0.68$ & RT, CW, planar antenna & \onlinecite{Chu2017a} \\\hline
         Collection rate & Tr:\textit{p}T & \SI{98}{Mph/s} & RT, CW, planar antenna & \onlinecite{Lee2011}\\\hline
         \multirow{3}{3cm}{Detection rate} & Tr:\textit{p}T & \SI{50}{Mcts/s} & RT, CW, planar antenna & \onlinecite{Lee2011}\\
         & DBATT:naph & \SI{4.4}{Mcts/s} & LT, CW, SIL & \onlinecite{Maser2017}\\
         & DBT:\textit{p}DCB & \SI{8}{Mcts/s} & LT, CW, plasmonic antenna + SIL & \onlinecite{Zirkelbach2020}
         \\\hline
         \multirow{2}{4cm}{Single-molecule indistinguishability, $V(0)$} & DBATT:\textit{n}-tetradecane & $0.93\pm0.05$ & LT, CW, atom vapor filter & \onlinecite{Rezai2018} \\
          &  DBT:Ac nanocrystals & $0.88\pm0.04$ & LT, pulsed and CW & \onlinecite{Duquennoy2022,Lombardi2021,Schofield2022b}
         \\\hline
         \multirow{3}{4cm}{Two-molecule indistinguishability, $V(0)$} & DBATT:\textit{n}-tetradecane & $\sim 0.3$ & LT, CW & \onlinecite{Lettow2010}\\
          & DBT:Ac nanocrystals & $0.4\pm0.07$ & LT, pulsed & \onlinecite{Duquennoy2022}\\
          & DBT:Ac cosublimated & $0.98\pm 0.02$ & LT, CW & \onlinecite{Huang2025}
        \\
    \end{tabular}
    \end{ruledtabular}
    \footnotesize
    \footnotetext[1]{RT: room temperature, LT: low temperature, CW: continuous wave excitation, pulsed: pulsed excitation}
\end{table*}

In table \ref{tab:figures_of_merit}, we summarize the properties of molecular single-photon sources using common figures of merit. The most important figure of merit is the purity $p$ of the source, which can be defined as the probability of finding no more than one photon at a given time. It can be directly calculated from the second-order correlation function $g^{(2)}(\tau)$ at zero time delay using the expression $p=1-g^{(2)}(0)/2$. For a pulsed source, this value is obtained by comparing the central area around zero time delay to the area under the periodic coincidence peaks at multiples of the repetition rate of the excitation laser \cite{Michler2000, Santori2004}. The main factors affecting the single-photon purity are the signal-to-background ratio in the emission and the presence of other emitters within the excitation and collection volume; therefore, it also often serves as a confirmation of the single-emitter nature of the observed optical signal.

The photon collection efficiency has already been introduced in the previous section. Related to this quantity is the photon collection rate, which quantifies the number of collected photons per second. This absolute number depends strongly on the excitation intensity and properties of the emitter itself, like its spontaneous emission rate. An alternative classification often used in pulsed operation is the so-called brightness, which states the number of collected photons by the first lens per excitation pulse \cite{Somaschi2016}.

The photon collection efficiency and collection rate are idealized quantities which do not include losses of the setup and the quantum efficiency of the detector. Therefore, one also often specifies the detection rate, which gives the true count rate at the detector, and the total efficiency (also called end-to-end efficiency), defined for a pulsed single-photon source as the probability to detect a photon after the emitter has been excited. These quantities are key parameters in the context of squeezing or linear optics quantum computation.

Finally, there is the indistinguishability of photons produced by the same emitter or by independent emitters. This property is important whenever two-photon interference is required \cite{Knill2001,Bouwmeester1997}, and it is usually characterized by the Hong-Ou-Mandel (HOM) visibility\cite{Santori2002,Rezai2018,Schofield2022b} $V(0)=(g_\perp^{(2)}(0)-g_\parallel^{(2)}(0))/g_\perp^{(2)}(0)$, where $g_\perp^{(2)}(\tau)$ and $g_\parallel^{(2)}(\tau)$ are the coincidence rates in the HOM experiment for the distinguishable and indistinguishable emission, respectively. Similar to the purity, its estimate for a pulsed source case is based on areas under the corresponding coincidence peaks.

Note that not all these properties of a single-photon source need to be optimized at the same time. In fact, the requirements on a source may differ significantly, largely depending on the application. For example, photon indistinguishability may not be needed for some applications like quantum imaging, while the requirement on the total efficiency is quite demanding \cite{Lounis2005}. 
 
\section{Challenges and outlook}
\label{sec:outlook}

\begin{figure*}
\includegraphics[width=16.5cm]{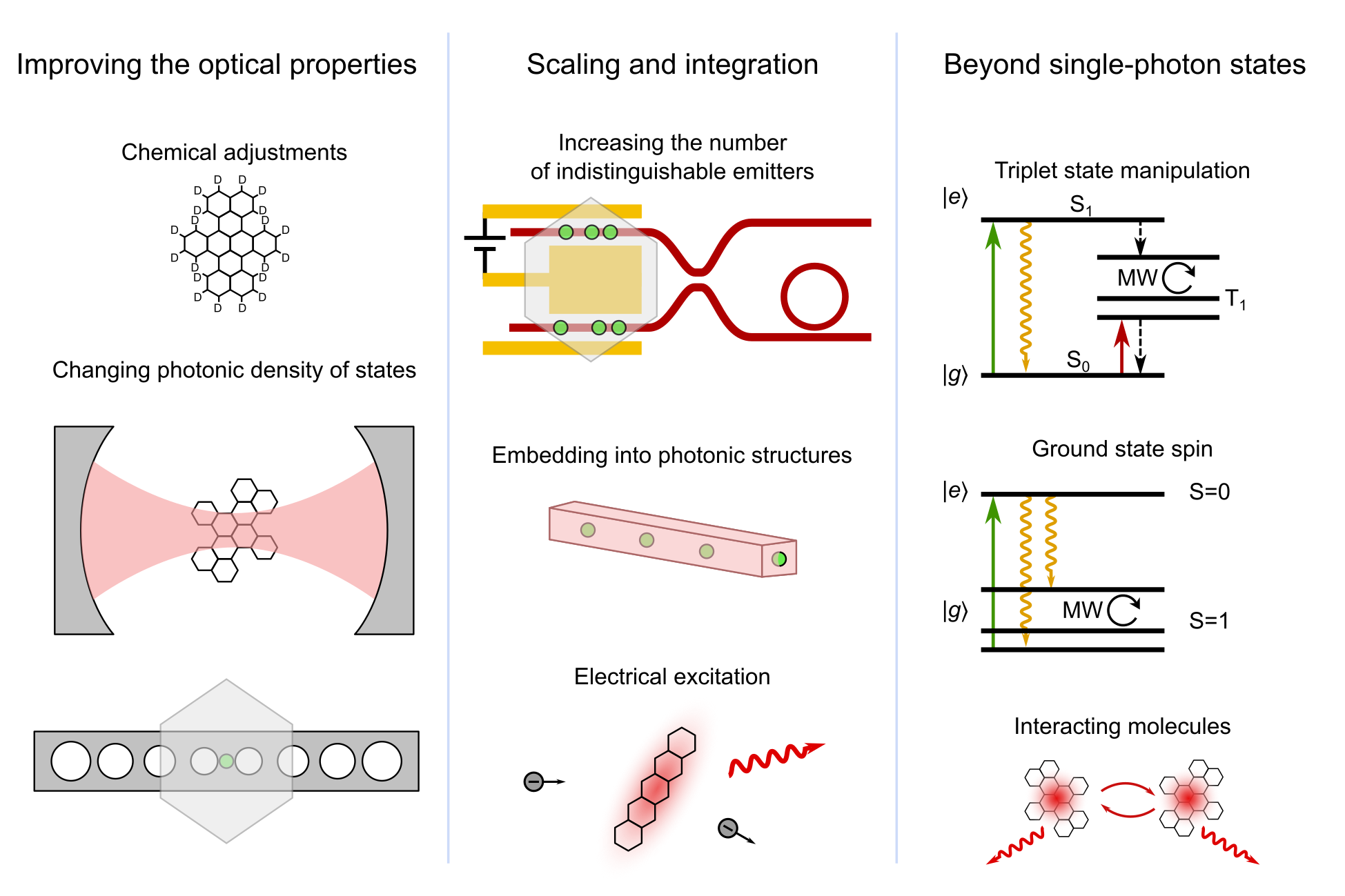}
\caption{\label{fig:outlook}\changed{Future directions for molecular single-photon sources. Left column: improving molecular properties such as emission efficiency and operation temperature range. This can be achieved by altering the guest molecule (e.g., deuteration for improved quantum efficiency), host system, or by selectively enhancing the ZPL transition with free-space Fabry-Perot resonators or monolithic structures such as photonic crystal cavities. Middle column: improving the integration and scaling the number of emitters. Currently, molecules have already been integrated into advanced nanophotonic circuits with integrated tuning electrodes and beam splitters for Hong-Ou-Mandel experiments. In the future, integration can be improved by inserting emitters directly into structures made out of appropriate host materials and by developing electrical excitation schemes. Right column: moving beyond single-photon emission. On an individual molecule level, this could be addressed by spin manipulation either in the long-lived triplet state or in the ground state of specially designed molecular emitters. Alternatively, several strongly coupled molecules can emit correlated or entangled photons.
}}
\end{figure*}

\changed{In this section, we discuss current challenges in the field of molecular SPSs and propose potential solutions. Furthermore, we highlight promising future directions (see Fig. \ref{fig:outlook}).}

\subsection{Emitter properties and collection efficiency}

Single dye molecules under the right conditions show a remarkable degree of stability and purity, producing highly indistinguishable photons for an unlimited time. However, like most solid-state quantum emitters, they still face the fundamental limitations of the efficiency at which indistinguishable photons can be generated and collected. Further improvements to the SPS performance demand that both the 00ZPL fraction $\eta_\mathrm{ZPL}$ and the collection efficiency $\eta_\mathrm{coll}$ are pushed as close to unity as possible.

In the case of molecules, the limitations of $\eta_\mathrm{ZPL}$ mainly arise from the 00ZPL branching ratio and quantum efficiency, both of which reduce the maximally achievable photon rate and limit the single photon probability in pulsed sources. There are two main strategies to address them. First is the improvement to the guest-host system. As discussed in section \ref{sec:emitter_parameters}, it has been recently discovered that the branching ratio can be strongly affected by the molecule environment, exceeding \SI{80}{\percent} in the best observed cases\cite{DeHaas2025}. This result provides a potential pathway to engineer the emitters to reach close to unity branching ratio. Similarly, a non-ideal quantum efficiency can be improved by deuteration of the guest and the host molecules\cite{Mishra2024}. In both of these cases, improvements in the computational models and the theoretical understanding of underlying processes will be a useful guide in the engineering of the material system.

The other, complementary, approach is the photonic engineering of the 00ZPL emission rate, as discussed in section \ref{sec:collection_methods}. A recent push towards incorporating resonant structures\cite{Wang2019,Rattenbacher2023,Lange2025} has already demonstrated significant modification of the branching ratio and the 00ZPL enhancement, although neither of these works has been directly optimized for single photon collection performance. Furthermore, molecules are readily compatible with other kinds of resonators such as monolithic Fabry-Perot\cite{Press2007} and bullseye\cite{Sapienza2015} cavities, which can strike a different balance between enhancement, stability, and ease of fabrication.

The collection efficiency improvement can also be achieved in several ways. On one hand, some room temperature experiments have already demonstrated remarkable collection efficiency of at least \SI{96}{\percent} and end-to-end efficiency of almost \SI{70}{\percent} using dielectric antennas\cite{Chu2017a}. Adapting this strategy to a cryogenic environment is challenging, but it could provide comparable performance while producing indistinguishable photons. Alternatively, this method can be combined with the cavity enhancement approach by optimizing the cavity geometry for the best outcoupling into the desired spatial mode.

On a more technical side, it is also beneficial to increase the temperature range over which the emitters display low dephasing and, correspondingly, high photon indistinguishability. The current experiments are typically performed at temperatures between \SI{1.4}{K}\cite{Rezai2018,Huang2025} and \SI{4.7}{K}\cite{Schofield2022b}, but it would be advantageous to further increase the operation temperature to accommodate more compact closed-cycle cryostats. Similarly to the 00ZPL branching ratio, one possibility lies in increasing the radiative decay rate of the emitter through cavity Purcell enhancement, which reduces the relative effect of extra dephasing; for example, the Purcell factor of $F\approx 40$ (enhanced linewidth of \SI{600}{MHz}) demonstrated in Ref. \onlinecite{Wang2019} would keep the emitter close to its lifetime limit for temperatures up to about \SI{10}{K}\cite{Clear2020}. The other way to minimize dephasing relies on a quantitative understanding and control of its origins. As discussed in section \ref{subsec:emitter_parameters_dephasing}, the temperature-induced dephasing arises from thermal phonons, and it is strongly dependent on the phonon interaction strength and density of states\cite{Skinner1988,Clear2020,Reitz2020}. This theoretical understanding, combined with the experimentally observed variation in the dephasing rate in different materials, suggests that it is possible to engineer guest and host systems to further reduce the effect of phonons. Recent experiments with surface molecules already demonstrate that the immediate neighborhood can have a strong effect on the molecules' vibronic spectra\cite{DeHaas2025}, suggesting that it is possible to engineer the molecule environment to affect its vibrational properties. At the same time, advances in modeling of interatomic potentials in molecular crystals\cite{Gurlek2025} should motivate more systematic numerical exploration of the local phonon environment and its effect on the emitter properties.

\subsection{Scaling and integration}

Another prominent goal is the integration of several independent sources of indistinguishable photons on a single chip. The most promising path here is using sub-wavelength waveguides \cite{Turschmann2017,Lombardi2018,Boissier2021,Ren2022,Huang2025} or monolithic cavities \cite{Rattenbacher2024,Lange2025} for photon collection, combined with micro-fabricated electrodes for Stark tuning\cite{Bauer2002,Lettow2010,Rattenbacher2024,Trebbia2022,Huang2025} or laser tuning\cite{Colautti2020,Duquennoy2022,Lange2024,Nobakht2025} of individual emitters to match their resonance frequencies. In this approach, the high achievable molecule density is a clear advantage, as it ensures that each waveguide or cavity can couple to tens or hundreds of emitters, each of which can be addressed individually via spectral or spatial selectivity, and tuned over a large range to achieve precise frequency matching.

A striking demonstration of this approach is given in a recent work, where two individually tunable lifetime-limited molecules were coupled to separate waveguides\cite{Huang2025}. The authors used resonant excitation and waveguide-based directional couplers to demonstrate on-chip Hong-Ou-Mandel interference, yielding a photon purity of $99\%$ and indistinguishability of $98\%$, while maintaining emitter-waveguide coupling above $10\%$. Importantly, the authors also demonstrated that each waveguide hosts dozens of molecules which can be tuned to the same resonance frequencies, clearly highlighting the scalability of the platform.

At the same time, molecular emitters have the unique advantage of not being strongly dependent on a particular host material, which allows for their integration into a wider range of systems. One underexplored possibility is utilizing self-assembled molecular cavities or waveguides\cite{Wang2014,Chen2021,Kim2025} instead of fabricated on chip-photonic devices. This approach can enable direct insertion of molecular emitters inside photonic structures and allow for larger-scale and more flexible production of nanoscopic photonic structures. \changed{Alternatively, incorporation into a stable transparent material which can be shaped using modern nanofabrication techniques would allow for placing the molecules directly into the nanophotonic waveguides or resonators, increasing the collection efficiency. Recent explorations of hBN\cite{Neumann2023,Farooqui2025} mentioned above represent the first step in this direction.}

The other important technical aspect for on-chip integration is the excitation method. The current experiments focus on optical methods thanks to their spectral selectivity (when 00ZPL excitation is used) and low implementation overhead. However, on-chip quantum applications would benefit from electrical excitation, as it can be integrated in a more flexible manner and produce less background compared to resonant excitation methods. Single-molecule electroluminescence has been previously demonstrated in STM-based plasmonic cavities and single-molecule junctions\cite{Zhang2017,Li2022}, including a PAH molecule (pentacene)\cite{Kong2021}. However, these techniques rely on the presence of metal electrodes in the immediate vicinity of molecules, which can induce dephasing or quenching, both detrimental to a high-quality SPS. Hence, novel methods, potentially inspired by organic light-emitting diode (OLED) technology, should be investigated in order to produce a deterministic electrically-driven single-photon source. An exciting possibility here is to combine molecular SPSs with the mature field of organic electronics \cite{Ostroverkhova2016,Liu2023,Li2024}, which should enable a more deterministic emitter placement (using, e.g., organic molecular beam epitaxy to implant emitters at specific depths), higher quality host materials, as well as the natural integration of electronic control.

\subsection{Generated photon states}

The last outstanding challenge we will discuss here relates to the quantum state of the emitted light. Some applications, especially the ones focusing on quantum computing, can require outputs more complicated than simple single-photon Fock states. Many types of solid-state emitters have a rich level structure with multiple long-lived states in the electronic ground state, which usually arise from an unpaired electron spin. These states can often be controlled using external microwave fields or special optical protocols, and thus have been successfully utilized for remote entanglement\cite{Ruf2021} or the generation of photonic cluster or graph states\cite{Cogan2023}. None of the PAHs used so far for single-photon generation have more than one long-lived ground state, which means that single isolated molecules were so far only used as single-photon emitters. 

There are several ways to extend molecular emitters in that direction. One is to use the long-lived triplet state as a quantum register or as an interface to nuclear spins\cite{Gurlek2024}. Here molecules with relatively long triplet state lifetime and stronger ISC, such as pentacene, are more promising due to more reliable initialization and readout of the triplet state. Indeed, already in the early days of single molecule spectroscopy, pentacene was used as a platform for single-molecule fluorescence-detected magnetic resonance (FDMR)\cite{Wrachtrup1993,Kohler1993}, which was later extended to demonstrating coupling of the triplet state energy to the hydrogen nuclear spin\cite{Kohler1999}. Some promising developments for spin manipulation and sensing using pentacene triplet state have also been demonstrated on ensembles at room temperature\cite{Mena2024,Singh2025,Singh2025b,Mann2025}\changed{, including experiments on surfaces\cite{Zheng2026}. While these studies rely on the triplet state excitation from the higher-energy excited singlet state via intersystem crossing (see Fig. \ref{fig:properties_illustrations}h), recent efforts have been dedicated to identifying direct optical transitions from the singlet ground state to the triplet excited state\cite{Smit2025}.}

The other opportunity is to design new molecular species with optically addressable ground-state spins. Recently, organometallic complexes\cite{Bayliss2020,Serrano2022,Shin2024} have demonstrated great promise in achieving optical spin initialization and readout, as well as relatively long spin coherence times of several microseconds\cite{Serrano2022} and the ability to tune spin properties by modifying the metal ions or ligand structure\cite{Bayliss2020,Laorenza2024}. \changed{Alternatively, purely organic carbene molecules have been shown to possess two unpaired electrons, resulting in a triplet ground state\cite{Roggors2025}. So far, all these experiments were performed on ensembles, but they clearly demonstrate a route towards single-molecule experiments.}

Finally, one can explore dipole-dipole coupling of several organic dye molecules\cite{Hettich2002,Trebbia2022,Lange2024}, which form collective sub-radiant and super-radiant states, effectively expanding the available state space. As this system of two emitters can produce a pair of photons closely spaced in time, it can serve as a basis for time-entangled photon emission. Moreover, as has been theoretically investigated\cite{Juan-Delgado2025}, under certain conditions, such a system of two coupled quantum emitters can produce polarization-entangled photon pairs. Such near-field coupling also brings further benefits, as the collective super-radiant state has an increased radiative decay rate, effectively playing the same role as a Purcell enhancement. It is important to note that this coupling does not have to rely on spatial proximity, but can also be realized through a shared photonic mode\cite{Nobakht2025}, which greatly increases the scalability potential.

\changed{It is notable, that a lot of the recent progress has been achieved through exploration of the vast chemical space of potential molecule candidates. These range from deuteration for improvements in quantum efficiency or spin environment\cite{Mishra2024,Roggors2025,Zheng2026} to atom or radical substitutions for control of optical or spin properties\cite{Bayliss2020,Laorenza2024,Mann2025} to investigation of new molecular classes\cite{Roggors2025,Feder2025,Chowdhury2025}. This, once again, highlights the potential of molecular engineering in improving and shaping molecular properties, especially when combined with modern computational and machine learning approaches\cite{Ohman2025}.
}

Manufacturing a truly on-demand indistinguishable single-photon source places strict demands both on the quantum emitter and on the surrounding optical system. Organic dye molecules are well-positioned to meet these challenges, with several different systems already demonstrating large tunability, near-unity quantum efficiency, high spectral stability, and negligible dephasing, resulting in high photon fluxes and indistinguishability. At the same time, their flexibility allows for integration with a variety of photonic systems, boosting the photon collection efficiency and bringing the molecules further towards the ideal two-level quantum emitter through Purcell enhancement. Recent efforts both in material system improvement and photonic engineering are on track to further boost the performance and versatility of molecular single-photon sources, keeping them competitive in the coming years.

\section{Acknowledgments}
The authors thank Vahid Sandoghdar for his continuous support. They also acknowledge financial support from the Max Planck Society, the Deutsche Forschungsgemeinschaft (DFG, German Research Foundation)—ID 429529648—TRR 306 QuCoLiMa (“Quantum Cooperativity of Light and Matter”) and the Free State of Bavaria via the Munich Quantum Valley light house project "QuMeCo".

\bibliography{aipmain}

\end{document}